\documentclass[a4paper,prd,showkeys,nofootinbib]{revtex4}

\usepackage[T1]{fontenc}
\usepackage[latin1]{inputenc}
\usepackage{graphicx}
\usepackage{amsmath}
\usepackage{amssymb}
\usepackage{slashed}
\DeclareMathOperator{\diag}{diag}

\begin{document}

\title{Spinors and the Weyl Tensor Classification in Six Dimensions}

\author{Carlos  Batista}
\email{carlosbatistas@df.ufpe.br}
\author{Bruno Carneiro da Cunha}
\email{bcunha@df.ufpe.br}
\affiliation{Departamento de F\'{\i}sica, Universidade Federal de Pernambuco, 50670-901
Recife - PE, Brazil}

\date{\today}

\begin{abstract}
A spinorial approach to 6-dimensional differential geometry is constructed and used to analyze tensor fields of low rank, with special attention to the Weyl tensor. We perform a study similar to the 4-dimensional case, making full use of the $SO(6)$ symmetry to uncover results not easily seen in the tensorial approach. Using spinors, we propose a classification of the Weyl tensor by reinterpreting it as a map from 3-vectors to 3-vectors. This classification is shown to be intimately related to the integrability of maximally isotropic subspaces, establishing a natural framework to generalize the Goldberg-Sachs theorem. We work in complexified spaces, showing that the results for any signature can be obtained by taking the desired real slice.

\end{abstract}

\maketitle
\section{Introduction}

 Despite its incontestable importance to Physics, spinors have had difficulties in being accepted as useful geometric tools in Physical theories, being replaced by the usual vector approach whenever possible. However there is a growing consensus that spinorial methods are very useful in a variety of field theory applications in the cases where there is \textit{a priori} knowledge that there is no mass gap developed \cite{WardWells}. Due to recent developments it became clear recently that spinor techniques can help fully use of the geometrical constraints, such as local Poincar\'{e} invariance \cite{scattering4d-1}. Building from work in generic solutions for the equations of motion for massless fields of arbitrary-spin, the perturbation theory for gauge fields can be restructured by enforcing Poincar\'{e} invariance rather than locality of interactions \cite{scattering4d-1}. This helps to reveal the integrability aspects of, for instance, maximally supersymmetric gauge theories, where one knows the renormalization group flow to be trivial. In general relativity spinors were particularly useful in four dimensions to construct explicit solutions of Einstein's equation, as well as clearly stating sufficient conditions for integrability. Despite all the successes, geometric applications of spinors in higher dimensions is still very much constrained to supersymmetric applications. Partly this is due to the inherent difficulty in dealing with generic dimensions consistently in spinorial language, even the basic definitions can be heavily dependent on the details of the symmetry group. In this paper we will focus on the six-dimensional case, covering the basic definitions and expanding on the relationship between differential geometry and algebraic structures. Nice reviews of spinorial applications in 4-dimensional differential geometry and field theory can be found in \cite{HuggettTod,WardWells}.


In the beginning of sixties Roger Penrose has started to use the spinor language in four-dimensional general relativity \cite{PenroseSpinor,Penrose}. This approach not only helped the search of new achievements in the field, but also it was of great importance to understand better previously known results. Causal structures are at the core of general relativity so it is natural that spinors prove to be very useful in this subject, specially in problems where the null directions are relevant.
Some examples in which spinorial techniques are much more elegant and concise than the usual tensor approach are the Petrov classification and the proof of the Goldberg-Sachs and the Mariot-Robinson theorems \cite{AdvancedGR,Penrose}. The intent of the present article is to follow some of these steps and introduce the spinor language in six-dimensional spaces, with the hope that this approach may clarify known results and pave way for further understanding. It will be shown that not only this clarification is achieved but also some new results are established here. In particular it will be shown that this language is very suitable when dealing with totally null structures, just as happens in four dimensions. Some previous material about spinors in six dimensions,
with field theoretical applications in mind,
can be found in \cite{Spinors in 6D}. General aspects of spinors in even dimensions were also used in \cite{Kerr-Robinson}, where the higher-dimensional Kerr theorem was investigated.

The Petrov classification is a scheme to classify the Weyl tensor in four dimensions that was of great importance to the geometrical theory of gravitation during the second half of last century. It can be used to judiciously find new solutions to the Einstein's equation, the main examples being the Kerr solution \cite{Kerr} and afterwards all other type $D$ vacuum solutions \cite{typeD - Kinnersley}. Of fundamental importance in the applicability of this classification was the Goldberg-Sachs (GS) theorem \cite{Goldberg-Sachs}, a theorem that relates the integrability of maximally isotropic submanifolds and the associated tangent bundle distribution in 4-dimensional manifolds to the algebraic form of the Weyl tensor \cite{Plebanski2,art2}. During the last decade many efforts have been made towards the creation of a higher-dimensional version of this classification as well as the GS theorem. In \cite{5D classification} a classification for the Weyl tensor in five dimensions was developed using spinor techniques and applications were made. A classification scheme valid in Lorentzian spaces of all dimensions was developed in \cite{CMPP}, dubbed the CMPP classification. Posterior work tried, with partial success, to generalize the GS theorem making connections between the optical matrix and the algebraic type of the Weyl tensor on this classification \cite{Durkee Reall,M. Ortaggio-GS theorem,M. Orataggio- GSII,M. Ortaggio-Robinson-Trautman}. Later a more geometrical approach towards a higher-dimensional generalization of the GS theorem was taken in \cite{HigherGSisotropic1,HigherGSisotropic2}, this path is based on the maximally isotropic distributions and will be of central importance here. Due to the relevance of these topics, the present work will deserve special attention towards the development of an useful classification for the Weyl tensor and the link between it and a generalized GS theorem. In six dimensions this was not attempted before using spinorial techniques, although this is probably the most suitable approach to deal with isotropic structures, at least in low dimensions.

In section \ref{Sec- Spinors R^6} it will be shown how the tensors of the vector space $\mathbb{C}\otimes\mathbb{R}^6\simeq\mathbb{C}^6$ can be expressed in terms of spinors. Particularly the totally null subspaces will be proven to have really simple representations in terms of spinorial subjects. Also a classification scheme for six-dimensional bivectors is developed. Section \ref{Sec- Reality Cond.} will show how the results obtained in the Euclidian space can be applied to the other signatures and the way to impose reality conditions in the spinor formalism will be displayed . After this, section \ref{Sec- Algeb. Classif.} deals with the issue of Weyl tensor algebraic classification from three perspectives: (i) it is studied the most simple forms that this tensor can take in the spinor approach, it turns out that these forms are too restrictive an should be of little use; (ii) the spinor representation is used to show that the Weyl tensor provides a natural map from self-dual 3-vectors to self-dual 3-vectors. Such map is used to define a classification scheme for the Weyl tensor and specially in the Euclidian signature this classification turns out be really simple, since in this case the map can be diagonalized; (iii) the well known CMPP classification will be expressed in terms of spinors and used to classify some previously defined types. Section \ref{Sec- Integrability Isotropic} will express the integrability condition for a maximally isotropic distribution, found in reference \cite{HigherGSisotropic2}, in an elegant and geometrical way in terms of pure spinors. It will also be shown that the integrability of such distributions is intimately related to the classification of the Weyl tensor based on the map in the 3-vector space. At the end of the section the generalized Mariot-Robinson theorem is briefly treated. Finally, section \ref{Sec- Examples} apply the results of this article to two important cases, the Schwarzschild and (a suitable analogue of) $pp$-wave 6-dimensional space-times. Appendix \ref{Appendix- Segre} presents a refinement of the Segre classification which will be suitable for our purposes, appendix \ref{Appendix-Spinor Basis} will show how to express specific basis of vectors, bivectors and 3-vectors in terms of spinors, while appendix \ref{Appendix - Clifford Alg.} will treat some details about the six-dimensional Clifford algebra using the index notation.

%

In this paper the spaces will be assumed to be complexified and it will be shown how the results on real spaces of arbitrary signature can be easily extracted from the complex case. Apart from the Lorentzian signature, the pure Euclidian case is of particular interest to string theory compactifications. The notion of algebraically special spaces provides a general geometric setting to study the phenomenon of reduced holonomy, and spinorial tools allow a direct correspondence to supersymmetry. Also, there is some attention drawn recently to the case of signature (4,2), which embeds ${\rm AdS}_5$ isometrically in six-dimensional flat space. There is hope that the tools developed here will be suited to deformations of the spaces mentioned above which still display enough algebraic structure to allow for analytical investigation.

\section{From $SO(6)$ to $SU(4)$}\label{Sec- Spinors R^6}
In this section we will present some relevant facts about the spinorial representation of the $SO(6;\mathbb{R})$ tensors. To this end we shall remember from the study of Clifford algebras that the universal covering group of $SO(6;\mathbb{R})$ is $SPin(\mathbb{R}^6)=SU(4)$ \cite{Lounesto}. Indeed, the latter is a double covering of the former, just as $SU(2)$ is a double covering for $SO(3;\mathbb{R})$. So every tensor of $SO(6;\mathbb{R})$ can be represented in terms of $SU(4)$ tensors. In an abuse of language, we will call the latter ``spinorial representations'' of the $SO(6;\mathbb{R})$ tensors.

The most basic representations of the group $SU(4)$ are the four-dimensional representations $\textbf{4}$ and $\overline{\textbf{4}}$ defined by\footnote{Throughout this article the following indices conventions will be used:
$A,B,C,\ldots$ are the spinorial indices and pertain to $\{1,2,3,4\}$; $\mu,\nu,\rho,\ldots$ are coordinate indices of $\mathbb{R}^6$, pertaining to $\{1,2,\ldots,6\}$; $a,b,c,\ldots$ are labels for a null frame of $\mathbb{C}\otimes\mathbb{R}^6$ and take the values $\{1,2,\ldots,6\}$; $i,j,k$ pertain to $\{1,2,3\}$; $r,s,t$ label a basis of (anti-)self-dual 3-vectors and runs from 1 to 10; $\varsigma,\upsilon,\kappa,\epsilon$ are four-dimensional spinor indices and can take the values 1 or 2; $p,q$ label a basis of Weyl spinors and pertain to $\{1,2,3,4\}$.}:

\begin{equation*}
    \textbf{4}:\;\; \zeta^{A}\stackrel{U}{\longrightarrow} U^A_{\phantom{A}B} \,\zeta^{B}\;\;\;\;\; ; \;\;\;\;\; \overline{\textbf{4}}:\;\; \gamma_{A}\stackrel{U}{\longrightarrow} \overline{U}_A^{\phantom{A}B}\, \gamma_{B}\;.
\end{equation*}
Where $U^A_{\phantom{A}B}$ is an unitary $4\times4$ matrix of unit determinant whose complex conjugate is $\overline{U}_A^{\phantom{A}B}$. A list of the low dimensional irreducible representations of $SU(4)$ can be found in \cite{Slansky}. From now on we shall call the 4-dimensional complex vectors $\zeta^{A}$ and $\gamma_{A}$ spinors. Note that if $\zeta^{A}$ transforms according to the representation $\textbf{4}$, then its complex conjugate,  $\overline{\zeta^{A}}$, will transform according to the representation $\overline{\textbf{4}}$. So it is natural to write $\overline{\zeta^{A}}=\overline{\zeta}_{A}$, that is, complex conjugation lowers the upper spinor indices and raises the lower spinor indices. It is also of fundamental importance to note that since $U^A_{\phantom{A}B}$ is an unitary matrix then the contraction of an upper index with a lower index is invariant by the group $SU(4)$:
$$ \zeta^{A}\gamma_{A}\;\;\stackrel{U}{\longrightarrow}\;\;\zeta^{A}\gamma_{A}\;,\; \textrm{Invariant by $SU(4)$}\,. $$
Defining $\varepsilon_{ABCD}$ to be the unique completely antisymmetric symbol such that $\varepsilon_{1234}=1$, it follows that the contraction of it with four arbitrary spinors, $\zeta^{A},\eta^{A},\varphi^{A}$ and $\xi^{A}$ is also invariant under $SU(4)$:
\begin{equation}\label{Epsilon contraction}
    \varepsilon_{ABCD}\zeta^{A}\eta^{B}\varphi^{C}\xi^{D}\;\stackrel{U}{\longrightarrow}\; \det(U)\, \varepsilon_{EFGH}\zeta^{E}\eta^{F}\varphi^{G}\xi^{H} = \varepsilon_{ABCD}\zeta^{A}\eta^{B}\varphi^{C}\xi^{D}.
\end{equation}

Since a vector of $SO(6;\mathbb{R})$, $V^\mu$, has 6 degrees of freedom it follows that its spinorial equivalent must transform in a six-dimensional representation of $SU(4)$. The group $SU(4)$ has two representations with this dimension, the representation $V^{AB}=-V^{BA}$, denoted by $\textbf{6}$, and representation $V_{AB}=-V_{BA}$, denoted by $\overline{\textbf{6}}$. But both representations are equivalent, since one can be transformed into the other by means of the symbol $\varepsilon_{ABCD}$. On the same vein, the bivectors of $SO(6;\mathbb{R})$, $B_{\mu\nu}=-B_{\nu\mu}$, have 15 degrees of freedom, so they must be in a 15-dimensional representation of $SU(4)$. The representation $\textbf{15}$ of $SU(4)$ is given by $B^A_{\phantom{A}B}$ with $B^A_{\phantom{A}A}=0$.

This last result about the bivectors establishes that it is possible to associate to every bivector an operator in the space of spinors with vanishing trace and whose action is $\chi^A\mapsto\chi'^A=B^A_{\phantom{A}B}\chi^B$. So the bivectors of $SO(6;\mathbb{R})$ can be classified according to the refined Segre type of the $4\times4$ matrix representation of this operator (see appendix \ref{Appendix- Segre}). As an example note that if $B^A_{\phantom{A}B}=\chi^A\gamma_B$ with $\chi^A\gamma_A=0$ then, in the notation of appendix \ref{Appendix- Segre}, the type of this bivector is $[|2,1,1]$. It should be observed that such classification would be quite hard and counterintuitive without spinors.

To find the spinorial equivalent of the 3-vectors of $SO(6;\mathbb{R})$, $T_{\mu\nu\rho}=T_{[\mu\nu\rho]}$, is a bit more involved, since 3-vectors have 20 degrees of freedom and there are a few ways to form a 20-dimensional representation of $SU(4)$. Once the 3-vectors are obtained from the linear combination of antisymmetric products of bivectors and vectors it follows from the above results that we should take a look at the product of the representations \textbf{15} and \textbf{6} of $SU(4)$. From \cite{Slansky} we see that $\textbf{15}\otimes\textbf{6}=\textbf{6}+\textbf{10}+\overline{\textbf{10}}+\textbf{64}$. So the 3-vectors must be in the representation  $\textbf{10}+\overline{\textbf{10}}$ of $SU(4)$, where the representation $\textbf{10}$ is given by $T^{AB}=T^{BA}$.

It will be of fundamental importance in this article to find the spinor equivalent of a tensor of $SO(6;\mathbb{R})$ with the symmetries of a Weyl tensor, $C_{\mu\nu\rho\sigma}=C_{[\mu\nu][\rho\sigma]}=C_{\rho\sigma\mu\nu}$, $C_{\mu[\nu\rho\sigma]}=0$ and $C^\mu_{\phantom{\mu}\nu\mu\sigma}=0$. A tensor with these symmetries has 84 degrees of freedom, which is the dimension of the representation $\Psi^{AB}_{\phantom{AB}CD}=\Psi^{(AB)}_{\phantom{AB}(CD)} $ with $\Psi^{AB}_{\phantom{AB}CB}=0$. There are other ways to find a 84-dimensional representations of $SU(4)$ but this is the only one that provides a natural map of bivectors into bivectors, just as the Weyl tensor does. More explicitly the equivalent of $B_{\mu\nu}\mapsto B'_{\mu\nu}=C_{\mu\nu\rho\sigma}B^{\rho\sigma}$ is $B^A_{\phantom{A}B}\mapsto B'^A_{\phantom{A}B}=\Psi^{AC}_{\phantom{AC}BD}B^D_{\phantom{D}C}$, note that $B'^A_{\phantom{A}A}=0$, as it should be for a bivector. So the spinorial equivalent of $C_{\mu\nu\rho\sigma}$ must be $\Psi^{AB}_{\phantom{AB}CD}$. Table \ref{Table spinors equivalent} sums up the equivalence relations between tensors and spinors.
\begin{table}
\begin{center}
\begin{tabular}{|c|c|c|}
  \hline
  $SO(6)$ Tensor & Spinorial Representation & Symmetries   \\ \hline
  $V^\mu$ & $V^{AB}$ & $V^{AB}=-V^{BA}$ \\ \hline
  $S_{\mu\nu}$ & $S^{AB}_{\phantom{AB}CD}$ & $S^{AB}_{\phantom{AB}CD}=S^{[AB]}_{\phantom{AB}[CD]}, \, S^{AB}_{\phantom{AB}CB}=0$ \\ \hline
  $B_{\mu\nu}$ & $B^A_{\phantom{A}B}$ & $B^A_{\phantom{A}A}=0$ \\ \hline
  $T_{\mu\nu\rho}$ & $(T^{AB},\widetilde{T}_{AB})$& $T^{AB}=T^{BA}, \widetilde{T}_{AB}=\widetilde{T}_{BA}$  \\ \hline
  $C_{\mu\nu\rho\sigma}$ & $\Psi^{AB}_{\phantom{AB}CD}$ & $\Psi^{AB}_{\phantom{AB}CD}=\Psi^{(AB)}_{\phantom{AB}(CD)},  \Psi^{AB}_{\phantom{AB}CB} = 0$ \\
  \hline
\end{tabular}
\caption{Spinorial equivalent of $SO(6;\mathbb{R})$ tensors. $V^\mu$ is a vector, $S_{\mu\nu}$ is a trace-less symmetric tensor, $B_{\mu\nu}$ is a 2-vector,\, $T_{\mu\nu\rho}$ is a 3-vector and $C_{\mu\nu\rho\sigma}$ is a tensor with the symmetries of a Weyl tensor.}\label{Table spinors equivalent}
\end{center}
\end{table}

Given two $SO(6)$ vectors, $V_1^\mu$ and $V_2^\mu$, it follows that the scalar product of them, $V_1^{\,\mu} V_{2\,\mu}$,  is the unique object, up to a multiplicative factor, invariant under $SO(6)$ and linear in $V_1$ and $V_2$. Now by equation (\ref{Epsilon contraction}) it follows that $\varepsilon_{ABCD}V_1^{\,AB} V_2^{\,CD}$ is invariant by $SU(4)$. Since this scalar is also linear in $V_1$ and $V_2$ then it must be proportional to the scalar product of these vectors. The proportionality constant can be arbitrarily chosen, in this article the choice will be $1/2$:
\begin{equation}\label{scalar product}
    V_1^{\,\mu} V_{2\,\mu} = \frac{1}{2} \varepsilon_{ABCD}V_1^{\,AB} V_2^{\,CD} = V_1^{\,AB} V_{2\,CD}\,.
\end{equation}
Where in the last equality it was defined that the lowering of an antisymmetric pair of spinorial indices is done by means of the contraction with $\frac{1}{2} \varepsilon_{ABCD}$. As an example of the use of $\varepsilon$ let us workout the spinorial representation of a trace-less symmetric tensor, $S_{\mu\nu} = S_{(\mu\nu)}$ and $S^\mu_{\phantom{\mu}\mu}=0$. This tensor has two indices, so it is obtained by the tensor product of two vector representations, $P^{AB\,CD}=P^{[AB]\,[CD]}$. A pair of antisymmetric indices can be lowered by means of $\frac{1}{2}\varepsilon$, yielding  $P^{AB}_{\phantom{AB}CD}=P^{[AB]}_{\phantom{AB}[CD]}$. This object can now be split into a scalar, $P^{AB}_{\phantom{AB}AB}$, a bivector, $(P^{AB}_{\phantom{AB}CB} -\frac{1}{4}\delta^A_CP^{DB}_{\phantom{DB}DB})$, and an object $S^{AB}_{\phantom{AB}CD}=S^{[AB]}_{\phantom{AB}[CD]}$ with zero trace, $S^{AB}_{\phantom{AB}CB}=0$. Since this last object has $20$ degrees of freedom it must be the spinor equivalent of tensors like $S_{\mu\nu}$. Now it is time to be more explicit in the relations between the tensors of $SO(6;\mathbb{R})$ and the tensors of $SU(4)$.

\subsection{The Precise Identifications}\label{SubSec Precise Ident.}
Just as in four dimensions the Infeld-van der Waerden symbols, $\sigma^{a}_{\;\kappa\dot{\epsilon}}$, establish a link between a vector and its spinorial representation, $v_{\kappa\dot{\epsilon}}=v_a\sigma^{a}_{\;\kappa\dot{\epsilon}}$, here, in six dimensions, there is an analogous symbol, denoted by $\Sigma_{\mu}^{\;AB}$. Given a vector of $\mathbb{C}\otimes \mathbb{R}^6$, $V^\mu$, its spinorial equivalent is defined by $V^{AB}=V^\mu\Sigma_{\mu}^{\;AB}$. Because of equation (\ref{scalar product}) these symbols must obey to the following relation.
$$g_{\mu\nu}=\frac{1}{2} \Sigma_{\mu}^{\;AB}\varepsilon_{ABCD}\Sigma_{\nu}^{\;CD}  $$
Where $g_{\mu\nu}$ is the metric of the vector space. It is useful to define a null frame for $\mathbb{C}\otimes \mathbb{R}^6$, $\{e_1,e_2,e_3,e_4=\theta^1,e_5 =\theta^2,e_6 =\theta^3\}$, defined to be such that the only non-zero inner products are $g(e_i,\theta^j)=\frac{1}{2}\delta_i^{\,j}$, in particular all base vectors are null. In this basis the symbols $\Sigma_{a}^{\;AB}$ can be given by
\begin{equation}\label{e^AB}
    \Sigma_{1}^{\;AB}=\delta_1^{[A}\delta_2^{B]}\;\; ; \;\; \Sigma_{2}^{\;AB}=\delta_1^{[A}\delta_3^{B]}\;\; ; \;\; \Sigma_{3}^{\;AB}=\delta_1^{[A}\delta_4^{B]}\;\; ; \;\; \Sigma_{4}^{\;AB}=\delta_3^{[A}\delta_4^{B]}\;\; ; \;\; \Sigma_{5}^{\;AB}=\delta_4^{[A}\delta_2^{B]}\;\; ; \;\; \Sigma_{6}^{\;AB}=\delta_2^{[A}\delta_3^{B]}\,.
\end{equation}
So given a vector $V=V^ae_a$ then its spinorial equivalent is $V^{AB}= V^a\Sigma_{a}^{\;AB}$. It is also immediate to obtain the spinorial representation of a trace-less symmetric tensor, $S^{AB}_{\phantom{AB}CD}=\Sigma_{a}^{\;AB}S^{ab}\Sigma_{b}^{\;EF}\frac{1}{2}\varepsilon_{EFCD}$. Now one can pose the question of how to convert the other tensors previously seen to the spinorial language. The answer to this must be worked case by case.

A bivector, $B_{\mu\nu}$, has two antisymmetric vectorial indices so that it must admit a spinorial representation $\mathfrak{B}^{AB\, CD}$ such that $\mathfrak{B}^{AB\, CD}=\mathfrak{B}^{[AB]\, [CD]}=-\mathfrak{B}^{CD\, AB}$. Thus, for example, if $B=(V_1\wedge V_2)$ then $\mathfrak{B}^{AB\, CD}=V_1^{\,AB}V_2^{\,CD}- V_2^{\,AB}V_1^{\,CD}$. Table \ref{Table spinors equivalent} imposes that $\mathfrak{B}^{AB\, CD}$ must be constructed in terms of $B^A_{\phantom{A}B}$. Up to an arbitrary scale factor there is only one way to do this:
\begin{equation}\label{B^AB CD}
  \mathfrak{B}^{AB\, CD} = B^{[A}_{\phantom{A}E}\varepsilon^{B]ECD}- B^{[C}_{\phantom{C}E}\varepsilon^{D]EAB}\,.
\end{equation}
Where $\varepsilon^{ABCD}$ is the completely antisymmetric symbol such that $\varepsilon^{1234}=1$. It is possible to invert the above relation and obtain $B^A_{\phantom{A}B}$ from $\mathfrak{B}^{AB\, CD}$ by a contraction with $\varepsilon_{ABCD}$:
\begin{equation}\label{B^AB from B^AB CD}
   B^A_{\phantom{A}B}=\frac{1}{4} \mathfrak{B}^{AC\, DE}\varepsilon_{CDEB}\,.
\end{equation}
Given a bivector $B_{\mu\nu}$ we can obtain its spinorial equivalent by means of the formula $\mathfrak{B}^{AB\, CD} = \Sigma_{a}^{\;AB}B^{ab}\Sigma_{b}^{\;CD}$. Subsequent application of equation (\ref{B^AB from B^AB CD}) then yields $B^A_{\phantom{A}B}$.

Now let us work out the 3-vectors. This kind of tensor may be used to map a vector into a bivector, $V^\mu\mapsto T_{\mu\nu\rho}V^\rho$, as well to map a bivector into a vector, $B_{\mu\nu}\mapsto T^{\mu\nu\rho}B_{\nu\rho}$. These maps must also be realizable in the spinorial language. An object that provides these maps in the spinorial language has the form $\tau^{ABC}_{\phantom{ABC}D} = \tau^{A[BC]}_{\phantom{A[BC]}D}$ with $\tau^{ABC}_{\phantom{ABC}A}=0$, since contracting it with the vector $V_{BC}$ produces a bivector, while contracting it with the bivector $B^D_{\phantom{D}A}$ results in a vector. From table \ref{Table spinors equivalent} it follows  that a 3-vector, $T_{\mu\nu\rho}$, is represented by the pair $(T^{AB},\widetilde{T}_{AB})$, so from this pair it must be possible to construct $\tau^{ABC}_{\phantom{ABC}D}$. Up to two arbitrary scale factors there is only one natural definition:
\begin{equation}\label{Tau ABC D}
    \tau^{ABC}_{\phantom{ABC}D} = T^{A[B}\delta_D^{C]} + \frac{1}{2} \widetilde{T}_{DE}\varepsilon^{EABC}\,.
\end{equation}
To know $\tau$ is equivalent to know the pair $(T,\widetilde{T})$, since the above relation can be easily inverted to give:
\begin{equation}\label{T from Tau}
    T^{AB} = \frac{2}{3} \tau^{ABC}_{\phantom{ABC}C}\;\;\;;\;\;\; \widetilde{T}_{AB}=\frac{1}{3} \varepsilon_{ACDE} \, \tau^{CDE}_{\phantom{CDE}B}\,.
\end{equation}
Since $T_{\mu\nu\rho}$ has three completely antisymmetric vector indices it must admit a spinorial representation of the form $\mathcal{T}^{AB\,CD\,EF}=\mathcal{T}^{[AB]\,[CD]\,[EF]}$ that is completely antisymmetric by the permutation between each pair of indices $AB$, $CD$ and $EF$. The object $\mathcal{T}$ should be constructed from $\tau$. Up to an arbitrary scale factor there is only one way to do this respecting the symmetries:
\begin{multline}\label{T^ABCDEF = tauEpsilon}
    \mathcal{T}^{AB\,CD\,EF}\;=\; \varepsilon^{GAB[C} \tau^{D]EF}_{\phantom{D]EF}G} \;+\; \varepsilon^{GCD[E} \tau^{F]AB}_{\phantom{F]AB}G} \;+\; \varepsilon^{GEF[A} \tau^{B]CD}_{\phantom{D]CD}G}\;+ \\ \;-\; \varepsilon^{GCD[A} \tau^{B]EF}_{\phantom{B]EF}G} \;-\; \varepsilon^{GEF[C} \tau^{D]AB}_{\phantom{D]AB}G} \;-\; \varepsilon^{GAB[E} \tau^{F]CD}_{\phantom{F]CD}G}\,.
\end{multline}
The inverse of this relation is:
\begin{equation}\label{Tau from T^AB CD EF}
    \tau^{ABC}_{\phantom{ABC}D} = \frac{1}{12} \mathcal{T}^{CB\,AE\,FG}\,\varepsilon_{EFGD}\,.
\end{equation}
Once given a 3-vector $T_{abc}$, its spinor equivalent $\mathcal{T}$ is obtained by contraction with the conversion symbols of equation (\ref{e^AB}), $\mathcal{T}^{AB\,CD\,EF} = \Sigma_{a}^{\;AB}\Sigma_{b}^{\;CD} \Sigma_{c}^{\;EF}\,T^{abc}$. Then equation (\ref{Tau from T^AB CD EF}) enables the calculation of $\tau$ and, eventually, equation (\ref{T from Tau}) yields the pair $(T,\widetilde{T})$.

Just as done above for bivectors and 3-vectors it is important to find the spinorial representation of the Weyl tensor that is obtained by means of the conversions symbols of equation (\ref{e^AB}). Given $C_{\mu\nu\rho\sigma}$ its spinor equivalent is $\mathcal{C}^{AB\,CD\,EF\,GH}=\Sigma_{a}^{\;AB}\Sigma_{b}^{\;CD}\Sigma_{c}^{\;EF}\Sigma_{d}^{\;GH}C^{abcd}$. This spinor equivalent must have the known symmetries of the Weyl tensor and, because of table \ref{Table spinors equivalent}, should be expressed in terms of $\Psi^{AB}_{\phantom{AB}CD}$. Up to an arbitrary scale there is only one way to do this respecting these symmetries:

\begin{equation}\label{C^AB CD EF GH}
    \mathcal{C}^{AB\,CD\,EF\,GH}= \Psi^{\mathring{A}\breve{E}}_{\phantom{AE}IJ}\varepsilon^{I\mathring{B}CD}\varepsilon^{J\breve{F}GH} +  \Psi^{\mathring{C}\breve{G}}_{\phantom{AE}IJ}\varepsilon^{I\mathring{D}AB}\varepsilon^{J\breve{H}EF} - \Psi^{\mathring{A}\breve{G}}_{\phantom{AE}IJ}\varepsilon^{I\mathring{B}CD}\varepsilon^{J\breve{H}EF} - \Psi^{\mathring{C}\breve{E}}_{\phantom{AE}IJ}\varepsilon^{I\mathring{D}AB}\varepsilon^{J\breve{F}GH}.
\end{equation}
Where, to simplify the notation, it was introduced the convention that two indices with $\mathring{}$ or with  $\breve{}$  should be anti-symmetrized, $F^{\mathring{A}\mathring{B}}= F^{AB}-F^{BA} = F^{\breve{A}\breve{B}} $. Laborious calculations show that $ \mathcal{C}^{AB\,CD\,EF\,GH}$ satisfy the expected constraints, $ \mathcal{C}^{AB\,CD\,EF}_{\phantom{AB\,CD\,EF}AB}=0$ and $\mathcal{C}^{AB\,CD\,EF\,GH} + \mathcal{C}^{AB\,EF\,GH\,CD} + \mathcal{C}^{AB\,GH\,CD\,EF}=0 $. The inverse of equation (\ref{C^AB CD EF GH}) is given by:
\begin{equation*}\label{Psi from C^AB CD EF GH}
    \Psi^{AB}_{\phantom{AB}CD} = 2^{-6}\; \mathcal{C}^{AI\,EF\,GH\,JB}\varepsilon_{IEFC}\,\varepsilon_{GHJD}\,.
\end{equation*}

It is also useful to know how the contraction of the Weyl tensor and a bivector is expressed in terms of spinors. Let $B'_{\mu\nu}=C_{\mu\nu\rho\sigma}B^{\rho\sigma}$. The immediate analogue of this equation in terms of spinors is $ \mathfrak{B}'^{AB\, CD} = \mathcal{C}^{AB\,CD\,EF\,GH}\mathfrak{B}_{EF\, GH}$. Now using equations (\ref{B^AB CD}) and (\ref{C^AB CD EF GH}) in the right hand side of the last equation and after this using relation (\ref{B^AB from B^AB CD}) we arrive at the important result:
\begin{equation}\label{C(B) by spinors}
    B'_{\mu\nu}=C_{\mu\nu\rho\sigma}B^{\rho\sigma}\;\;\;\Leftrightarrow\;\;\; B'^A_{\phantom{A}C}=-32 \Psi^{AB}_{\phantom{AB}CD}\,B^D_{\phantom{D}B}\,.
\end{equation}

With all these tolls at hand we can face the important task of finding how the isotropic structures are described in terms of spinors. It will be seen that the spinor language is the most suitable to deal with these mathematical objects.

\subsection{Isotropic Structures}\label{SubSec Isotropic structures}
A subspace of $\mathbb{C}\otimes \mathbb{R}^6$ is called isotropic or totally null when any vector belonging to it has zero norm. In this subsection it will be shown how the isotropic subspaces of one, two and three dimensions are expressed in terms of spinors. The final result will show that these subspaces have very simple spinorial representations.

Let $V^\mu$ be a vector tangent to a null direction of $\mathbb{C}\otimes \mathbb{R}^6$, $V^\mu V_\mu=0$. Now let us find the spinorial equivalent of the null vector $V$. Note that if $V^{AB}=\zeta^{[A}\eta^{B]}$ then because of (\ref{scalar product}) the norm of $V$ is zero. Conversely if $V$ is a null vector then it is possible to find spinors $\zeta$ and $\eta$ such that $V^{AB}=\zeta^{[A}\eta^{B]}$. To see this inductively suppose that $V^{AB}=\zeta^{[A}\eta^{B]}+\chi^{[A}\xi^{B]}$, then imposing that the norm of $V$ is zero by means of (\ref{scalar product}) we get $\varepsilon_{ABCD}\zeta^{A}\eta^{B}\chi^{C}\xi^{D}=0$, so that the spinors $\zeta,\eta,\chi$ and $\xi$ are not all linearly independent. Then, for example, $\xi=a\zeta+b\eta+c\chi$. Substituting this expansion in the definition of $V^{AB}$ we find $V^{AB}=(\zeta+b\chi)^{[A}(\eta-a\chi)^{B]}$. Thus can be concluded that:
\begin{equation}\label{null vector}
    V^{\mu}V_{\mu} = 0 \;\;\;\Leftrightarrow\;\; \exists \;\;\zeta^A,\eta^B\;\;|\;\; V^{AB}=\zeta^{[A}\eta^{B]}\,.
\end{equation}

  The simple bivector, $B^{\mu\nu}=V_1^{\,[\mu}V_2^{\,\nu]}$, is said to generate the 2-dimensional subspace spanned by the vectors $V_1$ and $V_2$. This bivector is called null if such subspace is isotropic, which means that $V_1^{\,\mu}V_{1\,\mu}=V_1^{\,\mu}V_{2\,\mu}=V_2^{\,\mu}V_{2\,\mu}=0$. Using equation (\ref{null vector}) it is simple matter to see that if $V_1$ and $V_2$ generate a totally null subspace then it is possible to find spinors $\zeta,\eta$ and $\chi$ such that $V_1^{\,AB}=\zeta^{[A}\eta^{B]}$ and $V_2^{\,AB}=\zeta^{[A}\chi^{B]}$. The spinorial representation of this null bivector is then $2\mathfrak{B}^{AB\, CD}=\zeta^{[A}\eta^{B]}\zeta^{[C}\chi^{D]}- \zeta^{[A}\chi^{B]}\zeta^{[C}\eta^{D]}$. Inserting this relation into equation (\ref{B^AB from B^AB CD}) we find that $B^A_{\phantom{A}B}= \frac{1}{8} \zeta^A(\varepsilon_{BCDE}\zeta^C\eta^D\chi^E)$. This result can be put in the following form:
  \begin{equation}\label{Null bivector}
    B_{\mu\nu} \; \;\textrm{is a null bivector} \;\;\; \Leftrightarrow \;\;\; \exists \;\;\zeta^A,\gamma_B\;\;\textrm{such that} \;\; \zeta^A \gamma_A=0\;\; |\;\;B^A_{\phantom{A}B}= \zeta^A \gamma_B\,.
  \end{equation}
   The isotropic subspace generated by the null bivector $B^A_{\phantom{A}B}= \zeta^A \gamma_B$ is the one spanned by the vectors of the form $V^{AB}=\zeta^{[A}\xi^{B]}$ with $\xi^{A}$ such that $\xi^{A}\gamma_A=0$.

The simple 3-vector $T^{\mu\nu\rho}=V_1^{\,[\mu}V_2^{\,\nu} V_2^{\,\rho]}$ is said to generate the 3-dimensional subspace spanned by the vectors $V_1$, $V_2$ and $V_3$. A 3-vector is called null if it is simple and the subspace generated by it is isotropic. In this case, by what was seen above, one of the options arise: (i) There exists spinors $\zeta^A$, $\eta^A$, $\chi^A$ and $\xi^A$ such that $V_1^{\,AB}=\zeta^{[A}\eta^{B]}$, $V_2^{\,AB}=\zeta^{[A}\chi^{B]}$ and $V_3^{\,AB}=\zeta^{[A}\xi^{B]}$; (ii)  There exists spinors $\zeta^A$, $\eta^A$ and $\chi^A$ such that $V_1^{\,AB}=\zeta^{[A}\eta^{B]}$, $V_2^{\,AB}=\zeta^{[A}\chi^{B]}$ and $V_3^{\,AB}=\eta^{[A}\chi^{B]}$. In the case (i) it can be shown that the spinorial representation of the 3-vector generated by these null vectors is $(T^{AB},\widetilde{T}_{AB}) \propto(\zeta^A\zeta^B,0)$, while in the case (ii) we have $(T^{AB},\widetilde{T}_{AB}) \propto(0,\gamma_A\gamma_B)$, with $\gamma_A=\varepsilon_{ABCD}\zeta^B\eta^C\chi^D$. So that we can conclude:
\begin{equation}\label{Null trivectors}
    T_{\mu\nu\rho}\;\;\textrm{is a null 3-vector} \;\;\;\Leftrightarrow \;\;\; (T^{AB},\widetilde{T}_{AB}) =  \begin{cases} (\zeta^A\zeta^B\;,\;0)\;\;\textrm{or} \\ (0\;,\;\gamma_A\gamma_B)\,. \end{cases}
\end{equation}
If $(T^{AB},\widetilde{T}_{AB}) =(\zeta^A\zeta^B,0)$ then the isotropic subspace generated by this 3-vector is the one spanned by the vectors $V^{AB}= \zeta^{[A}\xi^{B]}$ for all spinors $\xi^B$. While in the case $(T^{AB},\widetilde{T}_{AB}) =(0,\gamma_A\gamma_B)$ the isotropic subspace is the one spanned by vectors $V^{AB}=\varphi^{[A}\xi^{B]}$ for all $\varphi^A$ and $\xi^A$ such that $\varphi^A\gamma_A=0=\xi^A\gamma_A$. Note that in a complexified six-dimensional space the maximal dimension of an isotropic subspace is three. So 3-vectors like $(e_1\wedge e_2\wedge e_3)$ generate  maximally isotropic subspaces, these subspaces will be of fundamental importance throughout this article.

In general, not only in the null case, if the spinorial equivalent of a 3-vector is such that $\widetilde{T}_{AB}=0$ then the 3-vector is said to be self-dual, meaning that $\frac{1}{3!}\epsilon_{\mu\nu\rho\sigma\alpha\beta}T^{\sigma\alpha\beta}=iT_{\mu\nu\rho}$, where $\epsilon$ is the volume form of the vector space. Analogously, if $T^{AB}=0$ then the 3-vector is called anti-self-dual and obey to the equation  $\frac{1}{3!}\epsilon_{\mu\nu\rho\sigma\alpha\beta}T^{\sigma\alpha\beta}=-iT_{\mu\nu\rho}$.

One important message of this subsection is that the isotropic structures take really simple forms in the spinorial language. Note that the simplest form that a vector can take in the spinorial formalism is $V^{AB}=\chi^{[A}\eta^{B]}$ and this is exactly the form of a null vector. The simplest form that a bivector can take in the spinorial language is $B^A_{\phantom{A}B}=\chi^A\gamma_B$ with $\chi^A\gamma_A=0$, which is the form of a null bivector, that is a bivector ``tangent'' to a totally null plane. To finish, note that the simplest forms that a 3-vector can take in spinor formalism are just the ones that represent null 3-vectors.

\section{Reality Conditions and the Signatures }\label{Sec- Reality Cond.}
A general spinor $\xi^A$ has four complex degrees of freedom, which are acted from the left by the group $SU(4)$. So a general a vector $V^{AB}=V^{[AB]}$ constructed out of the spinor representation will have six complex degrees of freedom. If we want to take only the real vectors of $\mathbb{R}^6$ we will just have to impose the reality condition  $\overline{V^{AB}}=V_{AB}$. This condition makes sense from the group point of view because, as seen before, when the complex conjugate of a spinor is taken its upper indices must be lowered. Analogously a bivector is real if $\overline{B^A_{\phantom{A}B}}\equiv\overline{B}_A^{\phantom{A}B} = B^B_{\phantom{B}A}$. For 3-vectors the reality condition is $\overline{T^{AB}}\equiv \overline{T}_{AB} = \widetilde{T}_{AB}$, while the Weyl tensor is real if $\overline{\Psi^{AB}_{\phantom{AB}CD}} \equiv \overline{\Psi}_{AB}^{\phantom{AB}CD} = \Psi^{CD}_{\phantom{CD}AB}$.

For example, by equation (\ref{e^AB}) we see that the null vector $e_1$ has the spinorial representation $e_1^{\,AB}=\delta_1^{[A}\delta_2^{B]}$. Let us see if it is real. In one hand $\overline{e_1^{\,AB}}=\delta_1^{[A}\delta_2^{B]}= \delta^1_{[A}\delta^2_{B]}$, on the other $e_{1\,AB}\equiv\frac{1}{2}\varepsilon_{ABCD}e_1^{\,AB}= \delta^3_{[A}\delta^4_{B]}$. So the vector $e_1$ is not real, as could be predicted from the fact that in the Euclidian signature the only vector that is real and null is the zero vector.

\subsection{Changing the Signature}
Up to now only the Euclidian signature was considered, now we can go further and try to describe the six-dimensional spaces with other signatures in the spinor language. In this subsection it will be shown that most part of the above calculations can be carried to this more general situation. The only important difference is that while in the Euclidian signature it is easy to impose the reality condition using spinorial objects, in the other signatures this kind of operation is more laborious.

All the above results were extracted essentially from the fact that $SPin(\mathbb{R}^6)=SU(4)$. To analyze the six-dimensional spaces with other signatures, $\mathbb{R}^{5,1}$, $\mathbb{R}^{4,2}$ and  $\mathbb{R}^{3,3}$, it is useful to first study the space $\mathbb{C}^6$ and then choose conveniently the real slice in order to obtain the wanted signature \cite{Trautman,art1}. According to \cite{Trautman}, in the different signatures of a six-dimensional vector space we must have the following reality conditions on a null frame $\{e_i,\,e_{j+3}=\theta^j\}$ such that the only non-zero inner products are   $g(e_i,\theta^{j})=\frac{1}{2}\delta_{i}^{j}$:
\begin{equation}\label{Reality conditions signatures}
    \begin{cases}
    \mathbb{R}^6 \;\textrm{(Euclidian)}\rightarrow\;\;  \overline{e_1}=\theta^1\;,\; \overline{e_2}=\theta^2\;,\;\overline{e_3}=\theta^3 \\
    \\

    \mathbb{R}^{5,1} \;\textrm{(Lorentzian)}\rightarrow\;\;  \overline{e_1}=e_1\;,\; \overline{\theta^1}=\theta^1\;,\; \overline{e_2}=\theta^2 \;,\; \overline{e_3}=\theta^3\\
\\
    \mathbb{R}^{4,2} \rightarrow\;\;
    \begin{cases}
    \overline{e_1}=e_1\;,\; \overline{\theta^1}=\theta^1 \;,\; \overline{e_2}=e_2\;,\; \overline{\theta^2}=\theta^2 \;,\; \overline{e_3}=\theta^3 \\
        \overline{e_1}=-\theta^1 \;,\; \overline{e_2}=\theta^2 \;,\; \overline{e_3}=\theta^3
    \end{cases}\\
\\
    \mathbb{R}^{3,3} \;\textrm{(Split)}\rightarrow\;\;
    \begin{cases}
    \textrm{Real Basis}\\
    \overline{e_1}=e_1 \;,\; \overline{\theta^1}=\theta^1 \;,\; \overline{e_2}=-\theta^2 \;,\; \overline{e_3}=\theta^3\,.
    \end{cases}
        \end{cases}
\end{equation}

The isometry group of $\mathbb{C}^6$ is $SO(6;\mathbb{C})$, so that the group $SPin(\mathbb{C}^6)$ is the complexification of the group $SPin(\mathbb{R}^6)=SU(4)$, thus  $SPin(\mathbb{C}^6)=SL(4;\mathbb{C})$. So the tensors of $SO(6;\mathbb{C})$ can be expressed in terms of the representations of the group $SL(4;\mathbb{C})$. This last group has the following basic representations:
\begin{equation}\label{Represent. SL(4;C)}
    \textbf{4}:\;\; \zeta^{A}\stackrel{S}{\longrightarrow} S^A_{\phantom{A}B} \,\zeta^{B}\;\;\;\; ; \;\;\;\;
         \widetilde{\textbf{4}}:\;\; \widetilde{\gamma}_{A}\stackrel{S}{\longrightarrow} S^{-1\,B}_{\phantom{-1\,B}A} \,\widetilde{\gamma}_{B}\;\;\;\; ; \;\;\;\;
        \overline{\textbf{4}}:\;\; \gamma_{\dot{A}} \stackrel{S}{\longrightarrow} \overline{S}_{\dot{A}}^{\phantom{A}\dot{B}}\, \gamma_{\dot{B}} \;\;\; ; \;\;\;
        \widetilde{\overline{\textbf{4}}}:\;\; \widetilde{\zeta}^{\dot{A}}\stackrel{S}{\longrightarrow} \overline{S}^{-1\phantom{B}\dot{A}}_{\phantom{-1}\dot{B}} \,\widetilde{\zeta}^{\dot{B}}
\end{equation}
Where $S^A_{\phantom{A}B}$ is a complex $4\times4$ matrix with unit determinant, $S^{-1}$ is its inverse, $\overline{S}$ its complex conjugate and $\overline{S}^{-1}$ is the inverse of $\overline{S}$. The dot over some indices means that these indices are related to the complex conjugate of the fundamental representation. For the group $SU(4)$ the dot was not necessary because there the inverse of a group element is the transpose of the complex conjugate, so that the representations $\textbf{4}$ and $\widetilde{\overline{\textbf{4}}}$ are equivalent to each other, as well as the representations $\widetilde{\textbf{4}}$ and $\overline{\textbf{4}}$. But on the group $SL(4;\mathbb{C})$ such equivalences are not verified. Note also that the contractions $\zeta^A\widetilde{\gamma}_{A}$ and $\widetilde{\zeta}^{\dot{A}}\gamma_{\dot{A}}$ are invariant under the group $SL(4;\mathbb{C})$, but the contraction of a dotted index with an index without dot is not invariant.

 Except for the differences concerning the complex conjugation of spinors, almost everything seen before in the Euclidian signature carries to this more general situation. For example, the analogue of equation (\ref{Epsilon contraction}) is still valid, since $\det(S)=1$. Also the associations of table \ref{Table spinors equivalent} remain the same in the complex case, but now there is no natural way of imposing reality conditions in the spinor language. For example, $V^{AB}$ and $\overline{V^{AB}}= \overline{V}_{\dot{A}\dot{B}}$ can not be directly compared in the complexified case, since the former is in the representation $\textbf{6}$ of $SL(4;\mathbb{C})$, while the later is in the representation $\overline{\textbf{6}}$. If a charge conjugation operator is introduced then it is possible to make such comparison, but this path will not be taken here because it does not seem to play a role for the classification problem. Instead the imposition of reality conditions when the signature is not Euclidian will then be done in the tensor formalism, by means of equation (\ref{Reality conditions signatures}). It should be remembered that a similar phenomena happens in four dimensions, in which case it is easy to impose reality conditions when the signature is Lorentzian, but more difficult for the other signatures.

\section{Algebraic Classification of the Weyl Tensor}\label{Sec- Algeb. Classif.}
There are various ways of classifying a tensor field, depending on which use for the classification one has in mind. In four dimensions it is well known that the Petrov classification scheme is very useful and many solutions of Einstein's equation were found with the help of this classification, the Kerr solution being the most important example \cite{Kerr}. From the geometrical point of view this usefulness stems from the fact that in vacuum when the Weyl tensor is algebraically special according to the Petrov classification the manifold admits an integrable distribution of isotropic planes, this is the content of the Goldberg-Sachs theorem \cite{art2,Goldberg-Sachs,Plebanski2,Robinson Manifolds}. The intent of this section is to define classifications for the Weyl tensor in six dimensions that could be related to the integrability of isotropic structures, \textit{i.e.}, classifications that could be used to generalize, in some extent, to six dimensions the Goldberg-Sachs theorem.

This is an active area of research and there are two main paths being taken. The most popular approach is related to the well known CMPP classification \cite{CMPP,Coley_Review}, a scheme to classify tensors by means of the so called boost weights. Probably one of the biggest advantages of this path is that such classification is really simple and easy to understand. Some progress toward the generalization of the Goldberg-Sachs theorem using such classification was made in \cite{Durkee Reall,Pravda type D,M. Orataggio- GSII} and references therein, there the optical scalars are at the core of the treatment and the results are valid only for Lorentzian spaces. The second approach relates the algebraic form of the Weyl tensor to maximally isotropic structures \cite{HigherGSisotropic1,HigherGSisotropic2}, this is the path that will be taken here. Using this second classification Taghavi-Chabert was able to develop one interesting and promising generalization of Goldberg-Sachs theorem, valid for manifolds of arbitrary signature and dimension \cite{HigherGSisotropic2}. A short and nice review of some methods to classify the Weyl tensor in higher dimensions and its applications can be found in \cite{Reall - Review}, while a longer review focusing on the CMPP classification is available in \cite{Coley_Review}.

But none of the cited articles made use of spinors in six dimensions, so that this is the main originality of the present work. As proved in the preceding sections the spinorial language is the most suitable one to deal with isotropic structures, so that this article will shed light on useful facts that were not evident in the tensor formalism. In subsection \ref{SubSec Some Alg. Spec. Cases} the symmetries of $\Psi^{AB}_{\phantom{AB}CD}$ will be used to define some algebraically special forms to the Weyl tensor. Subsection \ref{SubSec 3-vectors MAP} will show that the Weyl tensor can be seen as an operator on the space of the 3-vectors, and this fact will be used to define a natural generalization of the Petrov classification. Finally, in \ref{SubSec Boost Weight} the boost weight classification will be described in terms of spinors.

From now on it will be assumed in this paper that the six-dimensional spaces treated so far are the tangent spaces of a six-dimensional manifold endowed with a non-degenerate metric $g$. Unless otherwise stated the tangent bundle will always be assumed to be complexified. The Weyl tensor is now the one associated to the Levi-Civita connection of the metric $g$ and all results are local.

\subsection{Some Algebraically Special Cases}\label{SubSec Some Alg. Spec. Cases}
In four dimensions the spinor equivalent of the Weyl tensor is the pair $(\Psi_{\varsigma\upsilon\kappa \epsilon}\,,\,\widetilde{\Psi}_{\dot{\varsigma}\dot{\upsilon}\dot{\kappa} \dot{\epsilon}})$. These objects are completely symmetric in its four indices, and since the indices can only take values 1 and 2 it follows that they can decomposed in terms of principal spinors. This is how the Petrov classification arises in the spinor approach \cite{Plebanski 1,AdvancedGR,Penrose}.

In six dimensions we can follow a similar path. Here the spinorial representation of the Weyl tensor is $ \Psi^{AB}_{\phantom{AB}CD}$, whose symmetries are $\Psi^{AB}_{\phantom{AB}CD}=\Psi^{(AB)}_{\phantom{AB}(CD)}$ and  $\Psi^{AB}_{\phantom{AB}CB} = 0$. Then some simple forms that are certainly algebraically special are defined in table \ref{Table Alg. Special}.
\begin{table}
\begin{center}
\begin{tabular}{ccc}
\hline
\;\; $(R,\widetilde{R})\rightarrow \Psi^{AB}_{\phantom{AB}CD} = \chi^A\chi^B\widetilde{\gamma}_C\widetilde{\gamma}_D\,; $ \; & \; $(R,\widetilde{S})\rightarrow \Psi^{AB}_{\phantom{AB}CD} = \chi^A\chi^B\widetilde{\gamma}_{(C}\widetilde{\xi}_{D)}$\,; \;&\; $(S,\widetilde{S})\rightarrow \Psi^{AB}_{\phantom{AB}CD} = \chi^{(A}\varphi^{B)}\widetilde{\gamma}_{(C}\widetilde{\xi}_{D)}$\,;\; \\
\end{tabular}\\
\vspace{\baselineskip}
\begin{tabular}{cc}
\;$(R,\widetilde{NS})\rightarrow \Psi^{AB}_{\phantom{AB}CD} = \chi^A\chi^B\widetilde{f}_{CD}$\,; \;&\;
$(S,\widetilde{NS})\rightarrow \Psi^{AB}_{\phantom{AB}CD} = \chi^{(A}\varphi^{B)}\widetilde{f}_{CD} $\,;  \\
\end{tabular}\\
\vspace{\baselineskip}
\begin{tabular}{cc}
\;$(R+\widetilde{R})\rightarrow \Psi^{AB}_{\phantom{AB}CD} = \chi^A\chi^B\widetilde{\gamma}_C\widetilde{\gamma}_D + \lambda  \gamma^A\gamma^B \widetilde{\chi}_C \widetilde{\chi}_D$\,; \;&\;
$(R \times \widetilde{R})\rightarrow \Psi^{AB}_{\phantom{AB}CD} = (\chi^A \chi^B + \varphi^A \varphi^B)(\widetilde{\gamma}_C \widetilde{\gamma}_D + \widetilde{\xi}_C \widetilde{\xi}_D)$.\;\\ \hline
\end{tabular}
\vspace{\baselineskip}
\caption{Some algebraically special types for the Weyl tensor. The labels comes from: Repeated(R), Simple(S) and Non-Simple(NS). Here $\chi^A\widetilde{\gamma}_A =\gamma^A\widetilde{\chi}_A =\chi^A\widetilde{\xi}_A = \varphi^A\widetilde{\gamma}_A= \varphi^A\widetilde{\xi}_A = \chi^A \widetilde{f}_{AB} = \varphi^A \widetilde{f}_{AB} = 0$, $\widetilde{f}_{AB} = \widetilde{f}_{BA}$ and $\chi^A\widetilde{\chi}_A=\gamma^A\widetilde{\gamma}_A= 1$. The types $(S,\widetilde{R})$, $(NS,\widetilde{R})$ and $(NS,\widetilde{S})$ can obviously be defined, as well as many other special types.}\label{Table Alg. Special}
\end{center}
\end{table}

Before proceeding it is important to note that if the Weyl tensor is real and the space has Euclidian signature then most of the special types defined in table \ref{Table Alg. Special} are prohibited. More precisely only type $(R+\widetilde{R})$ is allowed in this case. For example, suppose that the type is $(R,\widetilde{NS})$, then the reality condition $\overline{\Psi}_{AB}^{\phantom{AB}CD} = \Psi^{CD}_{\phantom{CD}AB}$ imposes that $\widetilde{f}_{AB} = \overline{\chi}_A \overline{\chi}_B $, so that $\Psi^{AB}_{\phantom{AB}CD} = \chi^A\chi^B\overline{\chi}_C \overline{\chi}_D $, but this is not compatible with the traceless condition $\Psi^{AB}_{\phantom{AB}CB} =0$, unless $\chi^A=0$. A similar phenomenon happens in four dimensions, where some algebraic types for the Weyl tensor are prohibited or allowed depending on the space signature \cite{art1}.

Now let us sum up some results concerning the algebraically special types defined on table \ref{Table Alg. Special}. To this end the tools of appendix \ref{Appendix-Spinor Basis} are used and the spinor basis is conveniently chosen. Below the non-zero components of the Weyl tensor will be displayed up to the trivial symmetries of this tensor.
\begin{itemize}
                                     \item Type $(R,\widetilde{R})$: \; $\Psi^{AB}_{\phantom{AB}CD} \propto \chi_1^{\,A}\chi_1^{\,B}\widetilde{\gamma}^2_{\,C}\widetilde{\gamma}^2_{\,D} \rightarrow$ The only non-zero component of the Weyl tensor is $C_{5656}$. Note that because of equation (\ref{Reality conditions signatures}) this type is not allowed in Lorentzian signature if the Weyl tensor is real, since in this case $\overline{C_{5656}}= C_{2323} =0$.
                                     \item Type $(R,\widetilde{S})$: \; $\Psi^{AB}_{\phantom{AB}CD} \propto \chi_1^{\,A}\chi_1^{\,B}\widetilde{\gamma}^2_{\,(C}\widetilde{\gamma}^3_{\,D)} \rightarrow$ The only non-zero component of the Weyl tensor is $C_{4656}$. If the Weyl tensor is real this type is not possible in the Lorentzian signature.
                                     \item Type $(S,\widetilde{S})$: \; $\Psi^{AB}_{\phantom{AB}CD} \propto \chi_1^{\,(A}\chi_2^{\,B)}\widetilde{\gamma}^3_{\,(C}\widetilde{\gamma}^4_{\,D)} \rightarrow$ The non-zero components of the Weyl tensor are $C_{5424} = -C_{6434}$. This type is realizable in the Lorentzian signature even if the Weyl tensor is real. In this case, using (\ref{Reality conditions signatures}) we have $\overline{C_{5424}} = C_{2454} = C_{5424}$ and $\overline{C_{6434}} = C_{3464} = C_{6434}$.
                                     \item Type $(R,\widetilde{NS})$: \; $\Psi^{AB}_{\phantom{AB}CD} = \chi_1^{\,A}\chi_1^{\,B}\widetilde{f}_{CD} \rightarrow$ The non-zero components of the Weyl tensor are $C_{4545}$, $C_{4546}$, $C_{4556}$, $C_{4646}$, $C_{4656}$ and $C_{5656}$. Also using equation (\ref{Reality conditions signatures}) it is easy to see that this type is not allowed when the signature is Lorentzian and the Weyl tensor is real.
                                     \item Type $(S,\widetilde{NS})$: \; $\Psi^{AB}_{\phantom{AB}CD} = \chi_1^{\,(A}\chi_2^{\,B)}\widetilde{f}_{CD} \rightarrow$ The non-zero components of the Weyl tensor are $C_{4543}$, $C_{4246}$ and $C_{4542}=-C_{4643}$. This type is permitted when the Weyl tensor is real and the signature is Lorentzian, in which case  $\overline{C_{4543}}=C_{4246}$, $\overline{C_{4542}}=C_{4542}$ and $\overline{C_{4643}} = C_{4643}$.
                                     \item Type $(R+\widetilde{R})$: \; $\Psi^{AB}_{\phantom{AB}CD} \propto (\chi_1^{\,A}\chi_1^{\,B}\widetilde{\gamma}^2_{\,C}\widetilde{\gamma}^2_{\,D} +\lambda \chi_2^{\,A}\chi_2^{\,B}\widetilde{\gamma}^1_{\,C}\widetilde{\gamma}^1_{\,D}) \rightarrow$  The non-zero components of the Weyl tensor are $C_{5656}$ and $C_{2323}=\lambda C_{5656}$. By equation (\ref{Reality conditions signatures}) if the Weyl tensor is real and the signature is Euclidian or Lorentzian then $\overline{C_{5656}}=C_{2323}=\lambda C_{5656}$, this implies that $|\lambda|=1$. So this type is allowed in the Euclidian and Lorentzian signatures for a real Weyl tensor if, and only if, $|\lambda|=1$.
                                     \item Type $(R \times \widetilde{R})$: \;  $\Psi^{AB}_{\phantom{AB}CD} \propto (\chi_1^{\,A}\chi_1^{\,B} + \alpha\, \chi_2^{\,A}\chi_2^{\,B})(\widetilde{\gamma}^3_{\,C}\widetilde{\gamma}^3_{\,D}  + \lambda\widetilde{\gamma}^4_{\,C}\widetilde{\gamma}^4_{\,D})\rightarrow $ The non-zero components of the Weyl tensor are $C_{6464}$, $C_{2424}=\lambda C_{6464}$, $C_{5454}=\alpha C_{6464}$ and $C_{3434} = \alpha\lambda C_{6464}$. If the Weyl tensor is real and the signature is Lorentzian then equation (\ref{Reality conditions signatures}) imposes that $|\lambda|=|\alpha|=1$. So this type is allowed in the Lorentzian signature for a real Weyl tensor only if $|\lambda|=|\alpha|=1$.
                                   \end{itemize}

The main conclusion that stems from the above results is that these algebraically special types are too special to be useful in general, since they impose that most of the Weyl tensor components vanish. Although very restrictive, in some specific situations these types may still have relevance. For example, these special types can be used to find new solutions to Einstein's equation.

\subsection{A Map from 3-vectors to 3-vectors}\label{SubSec 3-vectors MAP}
In its original form, Petrov classification have come from the analysis of the Weyl tensor as map from the space of bivectors to the space of bivectors \cite{Petrov,art1}. This map has the important property of sending (anti-)self-dual bivectors into (anti-)self-dual bivectors. This splits the Weyl operator that acts on the 6-dimensional space of bivectors into the direct sum of two operators that act in 3-dimensional spaces, $C'=C'^+ \oplus C'^-$, thus restricting enormously the possible algebraic types that this operator can have. Each of the two operators $C'^+$ and $C'^-$ will have only 6 possible types, the Petrov types. In particular, the null eigen-bivectors of the Weyl operator generate integrable isotropic planes when the Ricci tensor vanishes \cite{art2}, this is the geometrical content of the celebrated Goldberg-Sachs theorem. The intent of this section is to generalize to six dimensions this approach to the Petrov classification in a natural and geometrical way. This is important because it can lead to a geometric generalization of the Goldberg-Sachs theorem.

In six dimensions one cannot talk about a self-dual bivector, since the Hodge dual of a bivector is a 4-vector. But the 3-vectors can be self-dual or anti-self-dual, as already commented in subsection \ref{SubSec Isotropic structures}. Note also that, as proved in \cite{HigherGSisotropic2}, the maximally isotropic subspaces are at the core of the higher-dimensional generalization of the Goldberg-Sachs theorem. While in four dimensions these subspaces are generated by null bivectors, in six dimensions the maximally isotropic subspaces are generated by null 3-vectors. So it would be interesting if in six dimensions it was possible to see the Weyl operator as an operator in the space of 3-vectors. It would be of particular help if the spaces of self-dual and anti-self-dual 3-vectors were invariant by this operator, since this is the natural analog of what happens in four dimensions. Indeed we will see in this section that such a map exists and has the desired property.

In six dimensions the spinor equivalent of a 3-vector is the pair $(T^{AB},\widetilde{T}_{AB})$, while the spinor equivalent of the Weyl tensor is $\Psi^{AB}_{\phantom{AB}CD}$. Now it is immediate to see that we can associate to the Weyl tensor the following operator that sends 3-vectors into 3-vectors\footnote{In four dimensions the Weyl tensor is represented by $(\Psi_{\varsigma\upsilon\kappa \epsilon}\,,\,\widetilde{\Psi}_{\dot{\upsilon}\dot{\upsilon}\dot{\kappa} \dot{\epsilon}})$ and a bivector by $(\phi_{\alpha\beta}, \widetilde{\phi}_{\dot{\alpha}\dot{\beta}})$. The map of bivectors into bivectors in four dimensions is then $(\phi_{\varsigma\upsilon}, \widetilde{\phi}_{\dot{\varsigma}\dot{\upsilon}}) \mapsto (\Psi_{\phantom{\kappa \epsilon}\varsigma\upsilon}^ {\kappa \epsilon}\phi_{\kappa \epsilon}, \widetilde{\Psi}_{\phantom{\kappa \epsilon}\dot{\varsigma}\dot{\upsilon}}^ {\dot{\kappa }\dot{\epsilon}}\widetilde{\phi}_{\dot{\kappa}\dot{\epsilon}})$. The resemblance with six dimensions is a boon.}:
\begin{equation}\label{Psi 3-vec into 3-vec}
    C:\;\; (T^{AB},\widetilde{T}_{AB})\; \mapsto \; (T'^{AB},\widetilde{T'}_{AB}) = (\Psi^{AB}_{\phantom{AB}CD}T^{CD}\,,\, \Psi^{CD}_{\phantom{CD}AB}\widetilde{T}_{CD})\,.
\end{equation}
This map has the important property of preserving the subspaces of self-dual and anti-self-dual 3-vectors. To see this note that if $\widetilde{T}_{AB}=0$ then $\widetilde{T'}_{AB}=0$ and, analogously, if  $T^{AB}=0$ then $T'^{AB}=0$. This implies that we can write $C=C^+ \oplus C^-$, where $C^+$ gives zero when acts on anti-self-dual 3-vectors, while $C^-$ is trivial when acts on self-dual 3-vectors. So the $20\times20$ matrix that represents $C$ can be split into two blocks $10\times10$. The map $C^+$ is defined by $T^{AB} \mapsto T'^{AB} = \Psi^{AB}_{\phantom{AB}CD}T^{CD}$ and the operator $C^-$ has the action $\widetilde{T}_{AB} \mapsto \widetilde{T'}_{AB} =\Psi^{CD}_{\phantom{CD}AB}\widetilde{T}_{CD}$.

Up to now the resemblance between the map $C'$ in four dimensions and the operator $C$ in six dimensions was striking, but there exists one fundamental difference between the two cases. While in the former case the operators $C'^+$ and $C'^-$ are independent of each other (in the general complex case), in six dimensions the operators $C^+$ and $C^-$ have the same degrees of freedom. To see this explicitly let us make use of the 3-vector basis introduced in appendix \ref{Appendix-Spinor Basis}. If $C^{\pm}_{\;rs}$ are the matrices that represent the operators $C^{\pm}$ in the space of (anti-)self-dual 3-vectors then $C^+(T_s) = C^+_{\;rs}T_r$ and $C^-(\widetilde{T}^s) = C^-_{\;rs}\widetilde{T}^r$. Since $T_{r}^{\;AB}\widetilde{T}^s_{\;AB} = \delta^s_r$ then it follows that $C^{+}_{\;rs} = \widetilde{T}^r_{\;AB} \Psi^{AB}_{\phantom{AB}CD}T_{s}^{\;CD} = C^{-}_{\;sr}$. Note also that $C^{+}_{\;rr} = \Psi^{AB}_{\phantom{AB}AB}=0$, so the trace of $C^{\pm}$ is zero. Thus \emph{in six dimensions, for any signature, the operator $C^+$ is the transpose of $C^-$ and both operators have vanishing trace}. So while in four dimensions it makes sense to look for self-dual spaces(the ones with $C'^-=0$ and non-trivial $C'^+$) \cite{Plebanski3}, in six dimensions such concept cannot be introduced, since if $C^-=0$ then $C^+$ must also vanish.

The here proposed algebraic classification for the Weyl tensor amounts to compute of the refined Segre type of the operator $C^+$ (see appendix \ref{Appendix- Segre}). This operator is not arbitrary, in particular, as already observed, it must have zero trace. The traceless condition corresponds to only one of the 16 restrictions imposed by the identity $\Psi^{AB}_{\phantom{AB}CB}=0$. The remaining 15 conditions cannot be expressed in a form independent of the basis of 3-vectors, but certainly restrict the possible Segre types that this operator can have. This is an important point that needs further clarification. Note also that the algebraic type of $C^-$ is always the same of $C^+$, since one operator is the transpose of the other.

In the particular case of the Euclidian signature if the Weyl tensor is real then it is very simple to see what are the possible algebraic types for $C^+$. If in appendix \ref{Appendix-Spinor Basis} we choose the spinor basis to be $\chi_p^{\,A} = \delta_p^A$ then the complex conjugate, in this signature, of $T_{r}^{\;AB}$ is $\widetilde{T}^r_{\;AB}$ so that $\overline{C^{+}_{\;rs}} = \overline{\widetilde{T}^r_{\;AB} \Psi^{AB}_{\phantom{AB}CD}T_{s}^{\;CD}} = T_{r}^{\;AB} \Psi^{CD}_{\phantom{CD}AB} \widetilde{T}^s_{\;CD} = C^{+}_{\;sr}$. \emph{Thus if the Weyl tensor is real and the signature is Euclidian then the operator $C^+$ is Hermitean and, consequently, can be diagonalized}. Then the algebraic types of $C^+$ depends only on the 10 eigenvalues of this operator. Using the refined Segre classification defined in appendix \ref{Appendix- Segre} all numbers inside the square bracket will be one, remaining just the freedom of where to put the round brackets and of how many eigenvalues are zero.

As an example of the just defined classification note that if the Weyl tensor is type $(NS,NS)$ or more special,  $\Psi^{AB}_{\phantom{AB}CD}=f^{AB}\widetilde{g}_{CD}\neq0$,  then it is always possible to define a basis for the space of self-dual 3-vectors, $\{\mathbb{T}_r^{\;AB}\}$, such that $\mathbb{T}_r^{\;AB}\widetilde{g}_{AB}=\delta^2_r $ and $\mathbb{T}_1^{\;AB}=f^{AB}$. In this basis the operator $C^+$ has the following matrix representation:
\begin{equation*}\label{Matrix form C^+}
     [C^+] \,=\, \textrm{Blockdiag}(\left[
                     \begin{array}{cc}
                       0 & 1 \\
                       0 & 0 \\
                     \end{array}
                  \right], 0, 0, 0, 0, 0, 0, 0, 0)\,.
\end{equation*}
The refined Segre classification of this matrix is $[|2,1,1,1,1,1,1,1,1]$. In particular, all types defined in last subsection but $(R+\widetilde{R})$  are of the form $\Psi^{AB}_{\phantom{AB}CD}=f^{AB}\widetilde{g}_{CD}$, thus have this same algebraic classification. It is easy to see that the type $(R+\widetilde{R})$ with $\lambda\neq0$ admits a basis in which $[C^+] = \diag(\sqrt{\lambda}, -\sqrt{\lambda}, 0, \ldots, 0)$, so that the algebraic type is $[1,1|1,1,1,1,1,1,1,1]$.

Besides the approach used in the present section there are two other ways to represent the operator $C$. As seen in subsection \ref{SubSec Precise Ident.}, the pair $(T^{AB},\widetilde{T}_{AB})$ can be equivalently represented by an unique object, $\tau^{ABC}_{\phantom{ABC}D}$. The map $C$ is given in terms of this object by $\tau^{ABC}_{\phantom{ABC}D} \mapsto \tau'^{ABC}_{\phantom{ABC}D} = -3\Psi^{AE}_{\phantom{AE}DF} \tau^{FBC}_{\phantom{ABC}E}$. Note that $\tau'^{ABC}_{\phantom{ABC}D}$ has the necessary symmetries. Since this map deals only with objects that have tensorial equivalents, the Weyl tensor and the 3-vectors, then it must admit a tensor version. Indeed, in the tensorial formalism this map is proportional to the following map:
\begin{equation*}\label{C 3-vec into 3-vec}
    T_{\mu\nu\alpha}\,\mapsto \,T'_{\mu\nu\alpha}= C^{\rho\sigma}_{\phantom{\rho\sigma}[\mu\nu} \, T_{\alpha]\rho\sigma}\,.
\end{equation*}

Before moving on it is worth remembering that in six dimensions the Weyl tensor can also be seen as an operator on the space of bivectors whose action is $B_{\mu\nu}\mapsto B'_{\mu\nu}=C_{\mu\nu\alpha\beta}B^{\alpha\beta}$. We can algebraically classify this bivector operator using the Segre classification, just as we did with the operator $C$, and this can be used to refine and enhance other forms of classification. An extensive analysis of the bivector operator in higher dimensions was done in \cite{Bivector_Coley}. In spinorial language such map was given on equation (\ref{C(B) by spinors}) and hopefully the spinorial language could be used enlighten the study of the bivector operator in six dimensions.

\subsection{The Boost Weight Classification}\label{SubSec Boost Weight}
In this subsection the well known CMPP classification \cite{CMPP,Coley_Review} will be explained and expressed in terms of spinors. This is important because this is the simplest and most developed way to classify the Weyl tensor. This classification is most fruitful and plain in the Lorentzian case, therefore in the present subsection  this signature will be assumed, although at the end some comments will be made about other signatures.

Once introduced a null frame $\{e_i,\theta^j\}$, as defined before, let us define the boost transformation by:
\begin{equation}\label{Boost vectors}
    e_1\mapsto\lambda\, e_1\; ; \;\; \theta^1\mapsto\lambda^{-1}\,\theta^1 \; ; \;\;   e_2\mapsto e_2\; ; \;\; \theta^2\mapsto\theta^2 \; ; \;\; e_3\mapsto e_3\; ; \;\; \theta^3\mapsto\theta^3 \,,
\end{equation}
with $\lambda$ a real number. Note that equation (\ref{Reality conditions signatures}) says that in Lorentzian signature the vectors $e_1$ and $\theta^1$ are real, while the others are complex. This is the way to know which vectors of the null frame should be transformed by the boost.

A component of a tensor is said to have boost weight $b$ if under the above transformation it get multiplied by a factor of $\lambda^b$. For example, the component $T_{123}$ of the 3-vector $T_{\mu\nu\rho}$ has boost weight one, since $T_{123} = T_{\mu\nu\rho}e_1^{\;\mu}e_2^{\;\nu}e_3^{\;\rho} \stackrel{boost}{\longrightarrow} \lambda\, T_{\mu\nu\rho}e_1^{\;\mu}e_2^{\;\nu}e_3^{\;\rho}$. It is easy to see that the boost weights of the Weyl tensor components vary from 2 to $-2$. The CMPP classification then follows by the possibility of finding a null frame where the components of the Weyl tensor with certain $b$ vanish. Table \ref{Table CMPP types} defines the main CMPP types.
\begin{table}
\begin{center}
\begin{tabular}{|c|c|c|c|c|c|c|}
 \hline
CMPP Type & $O$ & $N$ & $III$ & $D$ & $II$ & $I$ \\
\hline
Vanishing Components & All & $b=2,1,0,-1$ & $b=2,1,0$ & $b=2,1,-1,-2$ & $b=2,1$ & $b=2$  \\
 \hline
\end{tabular}
\caption{Definition of the CMPP types. The second row says which components of the Weyl tensor should vanish according to the boost weight, $b$. For example, when the type is $I$ all components of boost weight two must vanish in some null frame.}\label{Table CMPP types}
\end{center}
\end{table}

If the null frame $\{e_i,\theta^j\}$ is such that all components of the Weyl tensor with boost weight $b=2$ vanish then the null vector $e_1$ is called a Weyl aligned null direction (WAND). In all dimensions this is equivalent to $e_1$ being a principal null direction according to the Bel-Debever criteria, $e_{1\,[\alpha}C_{\mu]\nu\rho[\sigma}e_{1\,\beta]}e_1^{\;\nu}e_1^{\;\rho}=0$ \cite{CMPP-Bel}. While in four dimensions there always exists a WAND, in higher dimensions this is not true. Moreover, in more than four dimensions it is possible to have a continuum of WANDs \cite{Reall - continuous WAND}, while in four dimensions at most four of these directions can exist. If the frame is such that the components of the Weyl tensor with $b=2$ and $b=1$ vanish, then the null vector $e_1$ is called a multiple WAND. In \cite{Durkee Reall} it was proved that a vacuum solution admits a multiple WAND if, and only if, there exists a multiple WAND that is geodesic. Another interesting property of this classification is that all known black hole solutions are type $D$.

In terms of the spinor basis defined in appendix \ref{Appendix-Spinor Basis}, the boost transformation, (\ref{Boost vectors}), is given by:
\begin{equation}\label{Spinors Boost}
    \begin{cases}
    \chi_{1}\mapsto \sqrt{\lambda} \,\chi_1 \;;\;\; \chi_2\mapsto \sqrt{\lambda} \,\chi_2 \;;\;\; \chi_3\mapsto \frac{1}{\sqrt{\lambda}}\, \chi_3 \;;\;\; \chi_4\mapsto \frac{1}{\sqrt{\lambda}}\, \chi_4 \\
    \widetilde{\gamma}^1 \mapsto  \frac{1}{\sqrt{\lambda}}\, \widetilde{\gamma}^1 \;;\;\; \widetilde{\gamma}^2 \mapsto  \frac{1}{\sqrt{\lambda}}\, \widetilde{\gamma}^2 \;;\;\; \widetilde{\gamma}^3 \mapsto  \sqrt{\lambda}\, \widetilde{\gamma}^3 \;;\;\; \widetilde{\gamma}^4 \mapsto  \sqrt{\lambda}\, \widetilde{\gamma}^4\,.
    \end{cases}
\end{equation}
With this at hand it is easy matter to know the boost weight of $\Psi^{AB}_{\phantom{AB}CD}$ components, next table summarizes this analysis.
\begin{table}[!htbp]
\begin{center}
\begin{tabular}{|c|c|c|c|c|c|c|c|c|c|}
\hline
\multicolumn{10}{| c |}{Components of $\Psi^{AB}_{\;\;\;\;\;CD}$ with boost weight $b$}\\
\hline
\; & $b=2$ & $b=1$ & $b=1$ & $b=0$ & $b=0$ & $b=0$ & $b=-1$ &$b=-1$ & $b=-2$ \\
\hline
$AB$ & $33,34,44$ & $13,14,23,24$ & $33,34,44$ & $13,14,23,24$ & $33,34,44$ & $11,12,22$ & $11,12,22$ & $13,14,23,24$ & $11,12,22$ \\
\hline
$CD$ & $11,12,22$ & $11,12,22$ & $13,14,23,24$ & $13,14,23,24$ & $33,34,44$ & $11,12,22$ & $13,14,23,24$ & $33,34,44$ & $33,34,44$ \\
\hline
\end{tabular}
\caption{The boost weight of the various components of $\Psi^{AB}_{\;\;\;\;\;CD}.$}\label{Table Psi AB_CD -boost}
\end{center}
\end{table}

As examples note that the special types defined in subsection \ref{SubSec Some Alg. Spec. Cases} are such that $(S,\widetilde{S})$, $(S,\widetilde{NS})$ and $(R\times\widetilde{R})$ are type $N$ in CMPP classification, $(R+\widetilde{R})$ and $(R,\widetilde{R})$ are type $D$, $(R,\widetilde{S})$ is type $III$ and $(R,\widetilde{NS})$ is type $II$. Remember that only types  $(S,\widetilde{S})$, $(S,\widetilde{NS})$, $(R\times\widetilde{R})$ and $(R+\widetilde{R})$ are allowed when the Weyl tensor is real and the signature is Lorentzian.

Commonly when working with the CMPP classification it is assumed that the signature is Lorentzian, this happens because in this case two non-orthogonal null directions, $e_1$ and $\theta^1$, can be distinguished in the null frame by the property of being real. But this classification can also be used in other non-Euclidian signatures \cite{ColeyPSEUD,Hervik-VSI}, moreover in \cite{art1} it was argued that it can be extended to Euclidian and complex spaces.

\section{Integrability of Maximally Isotropic Subspaces, the Goldberg-Sachs Theorem}\label{Sec- Integrability Isotropic}
The article \cite{HigherGSisotropic2} generalized in a geometrical way the important Goldberg-Sachs theorem to all dimensions greater than four. This generalization states that if the Weyl tensor satisfy certain algebraic conditions then the manifold admits an integrable distribution of maximally isotropic subspaces. The intent of the present section is to express these algebraic conditions in terms of spinors, this will prove to be very elegant and appropriate. It will also be shown that just as the four-dimensional Goldberg-Sachs theorem is intimately related to the map of bivectors into bivectors provided by the Weyl tensor \cite{art2}, the six-dimensional version of such theorem is connected to the map of 3-vectors into 3-vectors provided by the Weyl tensor. Before proceeding it is worth mentioning that very recently it was investigated in \cite{M. Orataggio- GSII} the existence of two-dimensional integrable  isotropic structures when the optical matrix is constrained and the Weyl tensor is type II on the CMPP classification. Integrable null distributions were also briefly investigated in \cite{Hervik-VSI}.

Let $N$ be a distribution of vector fields, over the complexified tangent bundle, that generates maximally isotropic subspaces at every point. The higher-dimensional version of the Goldberg-Sachs theorem says that in a Ricci-flat\footnote{Actually the theorem is more general and is also valid if certain components of the Ricci tensor are non-zero. Its original version is expressed in a conformally invariant way in terms of the Cotton-York tensor \cite{HigherGSisotropic2}. Here the vacuum condition is taken just for simplicity.} complex Riemannian manifold if the Weyl tensor is such that $C_{\mu\nu\rho\sigma}\, V_1^\mu V_2^\nu V_3^\rho = 0$ for all vectors fields $V_1$, $V_2$ and $V_3$ tangent to $N$ and generic otherwise\footnote{In \cite{HigherGSisotropic2} the proof given for this theorem requires that some generality conditions are satisfied by the Weyl tensor, so the imposition of ``generic otherwise'' is certainly sufficient, but it is not clear at all what is the necessary requirement. For example, in the section 3.4.2 of reference \cite{HigherGSisotropic1} some cases are shown in which the generality assumption can be relaxed. Also, at section 5.3 of \cite{M. Ortaggio-GS theorem} it is said that in five dimensions there exists many cases where such generality conditions can be neglected if the Ricci identity is used. As such, we will ignore this requirement in the present discussion.} then the distribution $N$ is locally integrable, \textit{i.e}, an involution under the Lie bracket. The converse of this theorem is not true, as exemplified in \cite{HigherGSisotropic2}. In six dimensions let us suppose that $N$ is generated by $\{e_1,e_2,e_3\}$ or, equivalently, the distribution is generated by the null 3-vector $(e_1\wedge e_2\wedge e_3)$, then we have:
\begin{equation}\label{Integrability condition}
  \textrm{Integrability Condition for $N$:}\;\;\;\; C_{ijka}\,=\,0 \;\;\;\; \forall \;\; i,j,k\;\in\;\{1,2,3\} \;;\;a\;\in\;\{1,2,\ldots,6\}\,.
\end{equation}

To express this condition in terms of spinors it is necessary to use table \ref{Table-WeylComp} of appendix \ref{Appendix-Spinor Basis}. Using this we get, for example, that $C_{12ja}=0$ is equivalent to the annihilation of  $\Psi^{4A}_{\phantom{4A}11}$ and  $\Psi^{4B}_{\phantom{4B}1C}$ for all $A$ and for all $B\neq1\neq C$. In general we have that:
\begin{equation}\label{Cijka = 0 ,v1}
   C_{ijka}\,=\,0\;\;\Leftrightarrow\;\; \begin{cases}
\Psi^{AB}_{\phantom{AB}CD}\chi_1^{\,C}\chi_1^{\,D}\widetilde{\eta}_B=0\\
\Psi^{AB}_{\phantom{AB}CD}\chi_1^{\,C}\widetilde{\xi}_A\widetilde{\eta}_B=0
\end{cases} \;\; \forall\;\;\widetilde{\xi},\, \widetilde{\eta}\;|\; \chi_1^{\,A}\widetilde{\xi}_A = \chi_1^{\,A}\widetilde{\eta}_A=0\,.
\end{equation}
In particular note that if we make $\chi^{A}=\chi_1^{\,A}$ in table \ref{Table Alg. Special} then, except for $(R+\widetilde{R})$, all types defined there  obey to the integrability condition $C_{ijka}=0$, since in these types we have $\Psi^{AB}_{\phantom{AB}CD}\chi^{C}=0$. Because of the identity $\Psi^{AB}_{\phantom{AB}CB}=0$ it follows that the first condition on the right hand side of equation (\ref{Cijka = 0 ,v1}) is a consequence of the second one, \textit{i.e.}, $C_{ijka}=0\Leftrightarrow\Psi^{AB}_{\phantom{AB}CD}\chi_1^{\,C}=0\;\;\forall \; A\neq1\neq B$. It is possible to workout a more elegant form to express this integrability condition using equation (\ref{EE Psi XXX}) of appendix \ref{Appendix-Spinor Basis} to arise at the following result:
\begin{equation}\label{Cijka = 0 ,v2}
  C_{ijka}\,=\,0\;\;\Leftrightarrow\;\; (\,\varepsilon_{AEFG}\,\varepsilon_{BHIJ}\,\Psi^{GJ}_{\phantom{GJ}CD}\,)\,\chi_1^{\,A}\, \chi_1^{\,B} \, \chi_1^{\,C}=0\,.
\end{equation}
Two other useful relations that can be proved are:
\begin{equation}\label{C_iabc}
 C_{ijab}\,=\,0\;\;\Leftrightarrow\;\; (\,\varepsilon_{AEFG}\,\Psi^{GJ}_{\phantom{GJ}CD}\,)\chi_1^{\,A}\, \chi_1^{\,C} =0 \;\;\;;\;\;\;C_{iabc}\,=\,0\;\;\Leftrightarrow\;\; (\,\varepsilon_{AEFG}\,\Psi^{GJ}_{\phantom{GJ}CD}\,)\chi_1^{\,A}\, =0\,.
\end{equation}

To understand in which way the above equations are elegant it is necessary to remember some important concepts about Clifford algebra and spinors. Given a spinor $\widehat{\psi}$, the null space associated to it, $N_{\widehat{\psi}}$,  is the vector space generated by the vectors that annihilate this spinor under the clifford action, $e\,\in\,N_{\widehat{\psi}}\,\Leftrightarrow\, \textbf{e}(\widehat{\psi})=0$ (see appendix \ref{Appendix - Clifford Alg.}). Because of the Clifford algebra, the vector space $N_{\widehat{\psi}}$ is always totally null. A spinor is called pure if its associated null space has the maximal dimension, such correspondence between maximally isotropic subspaces and pure spinors is one to one up to a scale.  It is well known that in even dimensions less or equal to six all Weyl spinors are pure, but in higher dimensions this is not true. Now using the conventions of appendices \ref{Appendix-Spinor Basis} and \ref{Appendix - Clifford Alg.} it is easy to see that the Weyl spinor of positive chirality $\chi_1^{\,A}$ is the pure spinor associated to the maximally isotropic vector subspace generated by $\{e_1,e_2,e_3\}$. So from what was said before it follows that \emph{if the Ricci tensor vanishes and the Weyl tensor satisfy the generality condition referred in \cite{HigherGSisotropic2} then}:
\begin{equation}\label{Integrab. Pure spinor}
 (\,\varepsilon_{AEFG}\,\varepsilon_{BHIJ}\,\Psi^{GJ}_{\phantom{GJ}CD}\,)\,\chi^{\,A}\, \chi^{\,B} \, \chi^{\,C}=0 \;\;\;\Rightarrow\;\;\; \textrm{The subspaces $N_{\widehat{\chi}}$ are locally integrable.}
\end{equation}
Where $N_{\widehat{\chi}}$ is the distribution of maximally isotropic subspaces associated to the pure spinor field $\chi^{\,A}$. So this equation expresses the integrability condition of totally null subspaces of maximal dimension as an algebraic condition involving the pure spinor field related to these subspaces and the spinorial equivalent of the Weyl tenor. This is very similar to what happens in four dimensions, where in vacuum the null planes associated to a pure spinor $\iota^\epsilon$ are integrable if, and only if, $\Psi_{\varsigma\upsilon\kappa\epsilon}\iota^\upsilon\iota^\kappa\iota^\epsilon=0$ \cite{HuggettTod}. Equation (\ref{Integrab. Pure spinor}) is valid only for the case when the pure spinor has positive chirality. In the case of a negative chirality spinor, $\widetilde{\chi}_{A}$, the algebraic condition on the left hand side of this equation must be substituted by
\begin{equation}\label{Integrab. Pure spinor NegChiral.}
   (\varepsilon^{AEFG}\,\varepsilon^{BHIJ}\,\Psi^{CD}_{\phantom{CD}GJ})\,\widetilde{\chi}_{\,A} \widetilde{\chi}_{\,B}  \widetilde{\chi}_{\,C}=0\,.
 \end{equation}

As an application of these integrability conditions note from table \ref{Table Psi AB_CD -boost} that, in a Lorentzian manifold, when the Weyl tensor  is type $N$ on the CMPP classification then $\Psi^{AB}_{\phantom{AB}CD}\chi_1^{\,C}=\Psi^{AB}_{\phantom{AB}CD}\chi_2^{\,C} = 0 $ and  $\Psi^{AB}_{\phantom{AB}CD} \widetilde{\gamma}^3_{\,A}=\Psi^{AB}_{\phantom{AB}CD}\widetilde{\gamma}^4_{\,A}$. In particular this implies, using equations (\ref{Integrab. Pure spinor}) and (\ref{Integrab. Pure spinor NegChiral.}), that the spinors $\chi_1$, $\chi_2$, $\widetilde{\gamma}^3$ and $\widetilde{\gamma}^4$ all obey to the integrability condition. The maximally isotropic distributions associated to these pure spinors are respectively $\{e_1,e_2,e_3\}$, $\{e_1,e_5,e_6\}$, $\{e_1,e_5,e_3\}$ and $\{e_1,e_2,e_6\}$. Although the integrability conditions are satisfied these distributions are not necessarily integrable, since the theorem of reference \cite{HigherGSisotropic2} assumes that the Ricci tensor obey to certain restrictions (Ricci-flat, for example) and that the Weyl tensor is generic, which are not true in general. It would be interesting to investigate what are the sharp necessary conditions that the Weyl tensor must obey in order to guarantee the integrability of these four distributions in vacuum type $N$ space-times.

Before moving on some comments are in order. The null subspaces associated to pure spinors of positive chirality are generated by self-dual 3-vectors, while the negative chirality spinors are associated to null subspaces generated by anti-self-dual 3-vectors. The difference between self-dual and anti-self-dual 3-vectors has no intrinsic meaning, since one type of 3-vector can be converted into the other by a simple change of sign in the volume form of the manifold. So in general we can assume that a null 3-vector is self-dual. Note also that although the calculations in this section are focused on the ``integrability of the 3-vector'' $(e_1\wedge e_2\wedge e_3)$ it must be kept in mind that given a null 3-vector it is always possible to choose a frame in which this 3-vector takes the form $(e_1\wedge e_2\wedge e_3)$. As an aside it is worth pointing out that the integrability condition for the existence of a complex structure on an Euclidian manifold is the existence of a pure spinor that obeys to the left-hand-side of equation (\ref{Integrab. Pure spinor}) and whose complex conjugate obeys to equation (\ref{Integrab. Pure spinor NegChiral.}).

\subsection{Integrability Condition and the Map from 3-vectors to 3-vectors - Complex Case}\label{SubSec Integ. Cond Complex}
The integrability condition of equation (\ref{Cijka = 0 ,v1}) can be equivalently expressed in terms of the properties of the operator $C^+$ that act on the space of self-dual 3-vectors. The first thing to note is that if $C_{ijka}=0$ then $T_1^{\;AB}= \chi_1^{\,A}\chi_1^{\,B}$ is an eigen-3-vector of the operator $C^+$. Indeed, if this integrability condition is satisfied then equation (\ref{Cijka = 0 ,v1}) implies $\Psi^{AB}_{\phantom{AB}CD}\chi_1^{\,C}\chi_1^{\,D}\widetilde{\gamma}^p_{\,B}= 0$ for $p\neq1$. This last equation is equivalent to the relation  $\Psi^{AB}_{\phantom{AB}CD}\chi_1^{\,C}\chi_1^{\,D} \propto  \chi_1^{\,A}\chi_1^{\,B}$. Now by means of equations (\ref{Tau ABC D}), (\ref{T^ABCDEF = tauEpsilon}) and (\ref{Null basis low indices}) it is possible to see that the self-dual 3-vector $T_1^{\;AB}= \chi_1^{\,A}\chi_1^{\,B}$ is proportional to $(e_1\wedge e_2\wedge e_3)$. So we arrived at the following result: \emph{The integrability condition of the maximally isotropic subspaces generated by the 3-vector $(e_1\wedge e_2\wedge e_3)$, $C_{ijka}=0$, implies that this 3-vector is an eigen-3-vector of the Weyl operator.} This generalizes the part of the Goldberg-Sachs theorem stating that in a four-dimensional vacuum manifold if a maximally isotropic distribution is integrable then the bivector that generates it is an eigen-bivector of the Weyl tensor \cite{art2}.

It is also easy to see that the second condition on the right hand side of equation (\ref{Cijka = 0 ,v1}) is equivalent to say that the subspace generated by the 3-vectors $\{T_1,T_2,T_3,T_4\}$ defined in appendix \ref{Appendix-Spinor Basis} is invariant by the operator $C^+$. So at the end we can conclude that \emph{the integrability condition $C_{ijka}=0$ is equivalent to imposing that the matrix representation of operator $C^+$ in the basis $\{T_r\}$ of equation (\ref{3-vector basis}) is of the following form}:
\begin{equation}\label{Cijka, [C^+] complex}
  [C^+] = \left[
            \begin{array}{cccccc}
              c_{11} & c_{12} & c_{13} & c_{14} & \ulcorner \phantom{4\times6} & \urcorner   \\
              0 & \ulcorner &    &\urcorner & \quad\quad4\times6 &   \\
              0 & & 3\times3 &  & \quad\quad \textrm{Block} &  \\
              0 & \llcorner &   & \lrcorner & \llcorner \phantom{4\times6}&   \lrcorner\\
              0 & 0 & 0 & 0 &  \ulcorner \phantom{4\times6}  & \urcorner   \\
              \vdots & \vdots & \vdots & \vdots & \quad\quad 6\times6&\\
              0 & 0 & 0 & 0 & \llcorner \phantom{4\times6} &\lrcorner \\
            \end{array}
          \right]
\end{equation}

\subsection{Integrability Condition and the Map from 3-vectors to 3-vectors - Euclidian Case}
Now let us suppose that the metric is real and with Euclidian signature, in particular the Weyl tensor is real. In this case it was proved in subsection \ref{SubSec 3-vectors MAP} that the matrix representation of operator $C^+$ in the basis $\{T_r\}$ is Hermitean. So, by using this property it follows that in this situation the integrability condition  $C_{ijka}=0$ is equivalent to say that the matrix representation of $C^+$ in the basis $\{T_r\}$ is the one of equation (\ref{Cijka, [C^+] complex}) with $c_{12}=c_{13}=c_{14}=0$ and with the $4\times6$ block vanishing. Also the $3\times3$ and $6\times6$ blocks must be Hermitean. Note that it is certainly much more difficult and less geometrical to derive this kind of result using the tensorial formalism instead of the spinor language. From the geometrical point of view the annihilation of more components of $C^+$ when the metric is real and the signature is Euclidian happens because, by equation ({\ref{Reality conditions signatures}}), if the subspace generated by $(e_1\wedge e_2\wedge e_3)$ is integrable then the subspace generated by $(\overline{e_1\wedge e_2\wedge e_3}) = (\theta^1\wedge\theta^2\wedge\theta^3)$ must also be integrable.

\subsection{Integrability Condition and the Map from 3-vectors to 3-vectors - Lorentzian Case}\label{SubSec- Lorentz}
In this subsection the metric is assumed to be real and Lorentzian, in particular the Weyl tensor is real. In this case the equation $C_{ijka}=0$ implies that $\overline{C_{ijka}}=C_{\overline{ijka}}= 0$. So the integrability condition for the subspaces generated by $(e_1\wedge e_2\wedge e_3)$ imposes the integrability condition for the subspaces generated by $(\overline{e_1\wedge e_2\wedge e_3}) = (e_1\wedge\theta^2\wedge\theta^3)$, where equation (\ref{Reality conditions signatures}) was used. The pure spinor associated to this last family of subspaces is $\chi_2^{\,A}$, as can be verified by means of equations (\ref{Null basis low indices}) and (\ref{Clifford algebra e_a}). So if $C_{ijka}=0$ then the right hand side of equation (\ref{Cijka = 0 ,v1}) is also valid if we put $\chi_2$ instead of $\chi_1$, thus the 3-vectors $T_1^{\;AB}=  \chi_1^{\,A}\chi_1^{\,B}$ and $T_5^{\;AB}=  \chi_2^{\,A}\chi_2^{\,B}$ are eigen-3-vectors of the operator $C^+$ and the following equations are valid:
\begin{equation}\label{Cijka, Lorentz}
    \Psi^{AB}_{\phantom{AB}CD}\chi_1^{\,C} = 0 \;\;\;, \quad\;\;\; \Psi^{EF}_{\phantom{EF}CD}\chi_2^{\,C}=0 \quad\quad \forall \;\; A\neq1\neq B \;,\; E\neq2\neq F.
\end{equation}
In particular these equations implies that $\Psi^{AB}_{\phantom{AB}CD}\chi_1^{\,C}\chi_2^{\,D} = 0$ for all $AB\neq1\,2$, that is, $T_2^{\;AB}= \sqrt{2}\,\chi_1^{\,(A}\chi_2^{\,B)}$ is an eigen-3-vector of $C^+$. In short, using the results of subsection \ref{SubSec Integ. Cond Complex}, we conclude that \emph{in Lorentzian signature if $C_{ijka}=0$ then the 3-vectors $T_1,T_2$ and $T_5$ are eigen-3-vectors of $C^+$ and the subspaces} $\textrm{Span}\{T_1,T_2,T_3,T_4\}$ \emph{and } $\textrm{Span}\{T_2,T_5,T_6,T_7\}$ \emph{are invariant under $C^+$.}

It is interesting to note that in Lorentzian signature the intersection of the maximally isotropic subspaces generated by $(e_1\wedge e_2\wedge e_3)$ and $(\overline{e_1\wedge e_2\wedge e_3})$ defines the real and null vector field $e_1$. Such vector field has special properties, for example, if $C_{ijka}=0$ it is easy to verify that $e_1$ is a Weyl Aligned Null Direction. Also, it is possible to prove that $e_1$ must be geodesic if these subspaces are integrable \cite{Robinson Manifolds}, which comes from the fact that the integral surfaces of an integrable maximally isotropic distribution are totally geodesic \cite{Mason-Chabert-KillingYano}. But, differently from what happens in four dimensions, in general the null congruence generated by $e_1$ is not shear-free\footnote{In higher dimensions the shear-free condition is much more restrictive than in four dimensions. One of the pioneering works to show this in an specific example was reference \cite{FrolovMyers5D}, where the 5-dimensional Myers-Perry black hole was studied and it was noted that the principal null
directions (according to the Bel-Debever criteria) are not shear-free. Thus proving that the Lorentzian Goldberg-Sachs theorem could not be trivially generalized to higher dimensions.}.

In vacuum four-dimensional Lorentzian spaces the Petrov type $D$ is characterized by the existence of four integrable totally null distributions of dimension two \cite{art2}. In order to generalize this concept, reference \cite{HigherGSisotropic1} suggests that a Lorentzian Ricci-flat manifold of dimension $d=2n$ should be said to have Petrov type $D$ if it admits $2^n$ maximally isotropic integrable distributions. Such definition is of relevance because it was established in \cite{Mason-Chabert-KillingYano}  a relation between the existence of a conformal Killing-Yano tensor and $2^n$ maximally isotropic integrable distributions. In six dimensions the Petrov type $D$ condition is equivalent to the existence of a null frame such that the distributions generated by the 3-vectors $(e_1\wedge e_2\wedge e_3)$, $(e_1\wedge \theta^2\wedge \theta^3)$, $(\theta^1\wedge e_2\wedge \theta^3)$, $(\theta^1\wedge \theta^2\wedge e_3)$,  $(\theta^1\wedge \theta^2\wedge \theta^3)$, $(\theta^1\wedge e_2\wedge e_3)$, $(e_1\wedge \theta^2\wedge e_3)$ and $(e_1\wedge e_2\wedge \theta^3)$ are all integrable. Since the pure spinors associated to these distributions are respectively $\chi_1^{\,A}$, $\chi_2^{\,A}$, $\chi_3^{\,A}$, $\chi_4^{\,A}$, $\widetilde{\gamma}^1_{\,A}$, $\widetilde{\gamma}^2_{\,A}$, $\widetilde{\gamma}^3_{\,A}$ and $\widetilde{\gamma}^4_{\,A}$ it follows that the integrability condition for these distributions are that spinors $\chi_1$, $\chi_2$, $\chi_3$ and $\chi_4$ obey to the left hand side of equation (\ref{Integrab. Pure spinor}) while spinors $\widetilde{\gamma}^1$, $\widetilde{\gamma}^2$, $\widetilde{\gamma}^3$ and $\widetilde{\gamma}^4$ obey to equation (\ref{Integrab. Pure spinor NegChiral.}). But it was proved in this subsection that if $\chi_1$ and $\chi_2$ obey equation (\ref{Integrab. Pure spinor}) then $T_1^{\;AB}= \chi_1^{\,A}\chi_1^{\,B}$, $T_2^{\;AB}= \sqrt{2}\,\chi_1^{\,(A}\chi_2^{\,B)}$ and $T_5^{\;AB}= \chi_2^{\,A}\chi_2^{\,B}$ are eigen-3-vectors of $C^+$. So if $\chi_1$, $\chi_2$, $\chi_3$ and $\chi_4$ obey to equation (\ref{Integrab. Pure spinor}) then all 3-vectors of the base defined in (\ref{3-vector basis}) are eigen-3-vectors of the Weyl operator. Thus we arrived at the result: \emph{In a Ricci-flat Lorentzian manifold if $C^+$ admits a diagonal matrix representation in a basis of the form defined in (\ref{3-vector basis}) then the Weyl tensor is of Petrov type $D$ \footnote{Ignoring the generality assumption of \cite{HigherGSisotropic2}.}.} It is interesting to note that it was not necessary to use the integrability conditions for the distributions generated by the anti-self-dual 3-vectors   $(\theta^1\wedge \theta^2\wedge \theta^3)$, $(\theta^1\wedge e_2\wedge e_3)$, $(e_1\wedge \theta^2\wedge e_3)$ and $(e_1\wedge e_2\wedge \theta^3)$ to establish this result.

\subsection{Generalized Mariot-Robinson Theorem}\label{SubSec Mariot-Robinson}
In a space of dimension $d=2n$ a simple $n$-form $\textbf{N}= V_{1\,\mu_1}\textbf{d}x^{\mu_1}\wedge\ldots\wedge V_{n\,\mu_n}\textbf{d}x^{\mu_n}$ is called null when the distribution generated by it, Span$\{V_1^{\,\mu_1},\ldots,V_n^{\,\mu_n}\}$, is isotropic. The generalized Mariot-Robinson theorem states that in $d=2n$ dimensions the maximally isotropic distribution generated by the null $n$-vector $\textbf{N}'\neq0$ is integrable if, and only if, there exists some function $h\neq0$ such that $\textbf{N}=h\textbf{N}'$ satisfies the equations $\textbf{d}\textbf{N}=0$ and $\textbf{d}(\star \textbf{N})=0$, where \textbf{d} is the exterior derivative and $\star\textbf{N}$ is the Hodge dual of $\textbf{N}$. This theorem was first proved in four-dimensional Lorentzian spaces in \cite{Robinson}, later it was extended to all signatures in \cite{McIntoshII} and finally it was generalized to all signatures and all even dimensions on reference \cite{Kerr-Robinson}, using spinor and twistor calculus.

By this theorem and by what was shown in preceding sections it follows that in a Ricci-flat six-dimensional manifold we can generally associate the existence of a null 3-form that is closed and co-closed to the existence of a positive chirality spinor field that obeys to the left hand side of equation (\ref{Integrab. Pure spinor}), or to the existence of a negative chirality spinor field obeying to equation (\ref{Integrab. Pure spinor NegChiral.}). In the former case the null $3$-form is self-dual, while in the later case the $3$-form is anti-self-dual. The generalized Mariot-Robinson theorem gives one more hint that while in four dimensions the bivectors are the relevant objects for the Weyl tensor classification and related matters, in manifolds of dimension $d=2n$ the $n$-forms are the important mathematical objects \cite{art4}.

\section{Examples}\label{Sec- Examples}
Now let us work out the topics treated in this article on two important examples for the general relativity theory, the Schwarzschild and a six-dimensional analogue of the $pp$-wave space-times.
\subsection{6D Schwarzschild}
The six-dimensional Schwarzschild space-time is the spherically symmetric vacuum solution of Einstein's equation whose metric is
$$ \textrm{ds}^2=-f^2\textrm{dt}^2 + f^{-2}\textrm{dr}^2 + r^2\{\textrm{d}\phi_1^2 + \sin^2\phi_1[\textrm{d}\phi_2^2 + \sin^2\phi_2\, ( \textrm{d}\phi_3^2 + \sin^2\phi_3\,\textrm{d}\phi_4^2 ) ]\}\,,$$
with $f^2=(1-\alpha r^{-3})$. A suitable null frame is defined by:
\begin{gather*}
  e_1=\frac{1}{2}\left(f\partial_r + f^{-1}\partial_t\right) \;\;;\;\; e_2=\frac{1}{2}\left(\frac{1}{r}\partial_{\phi_1} +\frac{i}{r\sin\phi_1}\partial_{\phi_2}\right) \;\;;\;\; e_3=\frac{1}{2}\left(\frac{1}{r\sin\phi_1\sin\phi_2}\partial_{\phi_3} +\frac{i}{r\sin\phi_1\sin\phi_2\sin\phi_3}\partial_{\phi_4}\right)  \;\;;\nonumber \\
  e_4=\frac{1}{2}\left(f\partial_r - f^{-1}\partial_t\right) \;\;;\;\; e_5=\frac{1}{2}\left(\frac{1}{r}\partial_{\phi_1} -\frac{i}{r\sin\phi_1}\partial_{\phi_2}\right) \;\;;\;\; e_6=\frac{1}{2}\left(\frac{1}{r\sin\phi_1\sin\phi_2}\partial_{\phi_3} -\frac{i}{r\sin\phi_1\sin\phi_2\sin\phi_3}\partial_{\phi_4}\right)\,.
\end{gather*}
With these definitions it is straightforward to prove the following commutation relations:
\begin{gather*}
  [e_1,e_2]=-\frac{f}{2r}e_2 \;\;;\;\; [e_1,e_3]=-\frac{f}{2r}e_3 \;\;;\;\; [e_1,e_4]=\frac{3\alpha}{4r^4}f^{-1}(e_1-e_4) \;\;; \\
  [e_2,e_3]=-\frac{1}{2r}(\cot\phi_1+i\frac{\cot\phi_2}{\sin\phi_1})e_3 \;\;;\;\; [e_2,e_4]=\frac{f}{2r}e_2 \;\;;\;\; [e_2,e_5]=\frac{\cot\phi_1}{2r}(e_2-e_5)  \;\;;\\ [e_2,e_6]=-\frac{1}{2r}(\cot\phi_1+i\frac{\cot\phi_2}{\sin\phi_1})e_6\;\;;\;\; [e_3,e_4]=\frac{f}{2r}e_3\;\;;\;\; [e_3,e_6]=\frac{\cot\phi_3}{2r\sin\phi_1\sin\phi_2}(e_3-e_6)\,.
\end{gather*}
The missing commutators can be obtained from the above equations using the following reality conditions: $\overline{e_1}=e_1$, $\overline{e_4}=e_4$, $\overline{e_2}=e_5$ and $\overline{e_3}=e_6$. From these commutation relations it is easily seen that there exists, at least, eight maximally isotropic integrable distributions in the six-dimensional Schwarzschild space-time, they are: $\{e_1,e_2,e_3\}$, $\{e_1,e_5,e_6\}$, $\{e_4,e_2,e_6\}$, $\{e_4,e_5,e_3\}$, $\{e_4,e_5,e_6\}$, $\{e_4,e_2,e_3\}$,  $\{e_1,e_5,e_3\}$ and $\{e_1,e_2,e_6\}$. So this space-time is of Petrov type D according to the definition of subsection \ref{SubSec- Lorentz}.

Up to the trivial symmetries the non-zero components of the Weyl tensor are:
$$C_{1414}=-\frac{3\alpha}{2r^5}\;;\;\; C_{1245}=C_{1346}=C_{1542}=C_{1643}=-\frac{3\alpha}{8r^5} \;;\;\; C_{2356}=C_{2552}=C_{2653}=C_{3636}=\frac{\alpha}{4r^5}\,.$$
From this it is easily concluded that the CMPP type of this space-time is $D$. Using table \ref{Table-WeylComp} of appendix \ref{Appendix-Spinor Basis} it can be shown that the spinor equivalent of the Weyl tensor in the six-dimensional Schwarzschild space-time is:
\begin{gather*}
  \Psi^{AB}_{\phantom{AB}CD} \,=\, -\frac{\alpha}{8r^5}[\chi_1^{\,A}\chi_1^{\,B}\widetilde{\gamma}^1_{\,C}\widetilde{\gamma}^1_{\,D} +\chi_2^{\,A}\chi_2^{\,B}\widetilde{\gamma}^2_{\,C}\widetilde{\gamma}^2_{\,D}
  + \chi_3^{\,A}\chi_3^{\,B}\widetilde{\gamma}^3_{\,C}\widetilde{\gamma}^3_{\,D}+
  \chi_4^{\,A}\chi_4^{\,B}\widetilde{\gamma}^4_{\,C}\widetilde{\gamma}^4_{\,D} ]\;+ \nonumber\\
  - 2\frac{\alpha}{8r^5}[ \chi_1^{\,(A}\chi_2^{\,B)}\widetilde{\gamma}^1_{\,(C}\widetilde{\gamma}^2_{\,D)} +
  \chi_3^{\,(A}\chi_4^{\,B)}\widetilde{\gamma}^3_{\,(C}\widetilde{\gamma}^4_{\,D)}   ] \;+ \nonumber\\
  +3\frac{\alpha}{8r^5}[ \chi_1^{\,(A}\chi_3^{\,B)}\widetilde{\gamma}^1_{\,(C}\widetilde{\gamma}^3_{\,D)} +
  \chi_1^{\,(A}\chi_4^{\,B)}\widetilde{\gamma}^1_{\,(C}\widetilde{\gamma}^4_{\,D)} +
  \chi_2^{\,(A}\chi_3^{\,B)}\widetilde{\gamma}^2_{\,(C}\widetilde{\gamma}^3_{\,D)} +
  \chi_2^{\,(A}\chi_4^{\,B)}\widetilde{\gamma}^2_{\,(C}\widetilde{\gamma}^4_{\,D)} ]\,.
\end{gather*}
Using this equation it is simple matter to prove that the spinors $\chi_1$, $\chi_2$, $\chi_3$ and $\chi_4$ obey to equation (\ref{Integrab. Pure spinor}), while spinors $\widetilde{\gamma}^1$, $\widetilde{\gamma}^2$, $\widetilde{\gamma}^3$ and $\widetilde{\gamma}^4$ obey to equation (\ref{Integrab. Pure spinor NegChiral.}), so that the integrability condition is satisfied by eight different pure spinor fields. These pure spinors are respectively the ones associated to the eight integrable maximally isotropic distributions listed above. Note also that any spinor of the form $\kappa=\chi_1+h\chi_2$, for any function $h$, obeys to the integrability condition of equation (\ref{Integrab. Pure spinor}), but the maximally isotropic distributions associated to these pure spinors, $\{e_1,(e_2+he_6),(e_3-he_5)\}$, are not integrable for $h\neq0$. This happens because the Weyl tensor of the Schwarzschild manifold does not obeys to the generality condition of Taghavi-Chabert \cite{HigherGSisotropic2}.

Actually these eight distributions are not the only integrable maximally isotropic distributions of the Schwarzschild space-time\footnote{The authors thank to Marcello Ortaggio for drawing our attention to this possibility. Comments on the same lines can also be found in section 8.3 of \cite{M. Orataggio- GSII}, where it was argued that Robinson-Trautman space-times with transverse spaces of constant curvature admit infinitely many isotropic structures. See also the footnote in the section 5.2 of reference \cite{M. Ortaggio-GS theorem}.}. Since the 4-sphere has a vanishing Weyl tensor it follows that it is conformally flat. Thus the Schwarzschild metric can be written as $\textrm{ds}^2=-f^2\textrm{dt}^2 + f^{-2}\textrm{dr}^2 + r^2g(y_j)[dy_1^2+ dy_2^2+ dy_3^2 + dy_4^2]$. Taking $n_1=a^i\partial_{y_i}$ and $n_2=b^j\partial_{y_j}$ with $a^i$ and $b^j$ constants it is simple matter to note that there are infinitely many choices of $a^i$ and $b^i$ that make $\{n_1,n_2\}$ an isotropic distribution. It is also immediate to note that such distributions are integrable, so that the distributions $\{e_1,n_1,n_2\}$ and $\{e_4,n_1,n_2\}$ are also integrable. Thus arriving at the result that there exist infinitely many integrable maximally isotropic distributions on this space-time.

Using the above expression for $\Psi^{AB}_{\phantom{AB}CD}$ it is easy to show that the matrix representation of the self-dual part of the Weyl operator in the basis $\{T_r\}$ defined on equation (\ref{3-vector basis}) is:
$$ [C^+]\,=\, -\frac{\alpha}{16r^5}\, \textrm{diag}(2,2,-3,-3,2,-3,-3,2,2,2)\,.$$
So that the refined Segre classification of this operator is $[(1,1,1,1,1,1),(1,1,1,1)|]$.

\subsection{$pp$-Wave}
Let us consider now the six-dimensional Kerr-Schild metric $g_{\mu\nu}=\eta_{\mu\nu}+ 2H k_\mu k_\nu$ with, $\eta_{\mu\nu}$ the metric of the six dimensional Minkowski space-time and $k$ a null vector field with respect to $\eta_{\mu\nu}$ such that $\partial_\mu k_\nu=0$ (the possible position dependence is chosen to be entirely in $H$). Then assuming that $k(H)=k^\mu\partial_\mu H=0$ it follows that the Ricci and the Riemann tensors are respectively given by:
$$ R_{\mu\nu}\,=\, -(\eta^{\alpha\beta}\partial_\alpha\partial_\beta H)k_\mu k_\nu \quad\quad;\quad\quad  R_{\mu\nu\rho\sigma}=2k_\sigma k_{[\mu} \partial_{\nu]} \partial_\rho H \,-\,2k_\rho k_{[\mu} \partial_{\nu]} \partial_\sigma H  \,. $$
Where $\eta^{\mu\nu}$ is the inverse of $\eta_{\mu\nu}$. If $\{x^0,x^1,x^2,x^3,x^4,x^5\}$ are cartesian coordinates for the Minkowski metric then defining the null vector field $k$ to be $k=\frac{1}{\sqrt{2}}(\partial_{x^0}+\partial_{x^1})$ and defining the null coordinates $u=x^0-x^1$, $v=x^0+x^1$, $z=x^2+ix^3$ and $w=x^4+ix^5$ then it follows that the metric of this space is given by:
$$\textrm{ds}^2\,=\, H\textrm{d}u^2 - \textrm{d}u\textrm{d}v+ \textrm{d}z\textrm{d}\overline{z} + \textrm{d}w\textrm{d}\overline{w} $$
A natural null frame to define is:
 \begin{equation}\label{nullFppwave}
 e_1=\partial_v\;;\;\; e_2=\partial_z\;;\;\; e_3=\partial_w\;;\;\;  e_4=-(\partial_u +H\partial_v)\;;\;\; e_5=\partial_{\overline{z}}\;;\;\;e_6=\partial_{\overline{w}}\,.
 \end{equation}
It is immediate to verify that the maximally isotropic distributions $\{e_1,e_2,e_3\}$, $\{e_1,e_5,e_6\}$, $\{e_1,e_5,e_3\}$ and $\{e_1,e_2,e_6\}$ are all integrable, since the relevant commutators vanish. Note also that the vector $k=\sqrt{2}e_1$ is covariantly constant so that the space-time is $pp$-wave and Kundt \cite{Kerr-Schild}. Now, for reasons of simplification, let us assume that $\partial_uH=\partial_wH=\partial_{\overline{w}}H =\partial_z \partial_{\overline{z}}H=0$, in which case the Ricci tensor vanishes and the only non-zero components of the Weyl tensor on the null frame of equation (\ref{nullFppwave}) are:
$$ C_{2424}=-\frac{1}{2}\partial_z \partial_zH \quad \quad ; \quad \quad C_{5454}=-\frac{1}{2}\partial_{\overline{z}} \partial_{\overline{z}}H\,.$$
Then the CMPP type of this Weyl tensor is $N$. From table \ref{Table-WeylComp} we see that $C_{2424}=4\Psi^{22}_{\phantom{22}33}$ and $C_{5454}=4\Psi^{11}_{\phantom{11}44}$, therefore the spinorial equivalent of the Weyl tensor is:
$$\Psi^{AB}_{\phantom{AB}CD} \,=\,\Theta\,\chi_1^{\,A}\chi_1^{\,B} \widetilde{\gamma}^4_{\,C}\widetilde{\gamma}^4_{\,D} \,+\, \widetilde{\Theta}\, \chi_2^{\,A}\chi_2^{\,B} \widetilde{\gamma}^3_{\,C}\widetilde{\gamma}^3_{\,D} \,.  $$
Where $\Theta=-\frac{1}{8}\partial_{\overline{z}} \partial_{\overline{z}}H $ and $\widetilde{\Theta}=-\frac{1}{8}\partial_z \partial_zH$. Since $\Psi^{AB}_{\phantom{AB}CD}\chi_1^{\,C}=\Psi^{AB}_{\phantom{AB}CD}\chi_2^{\,C}=0$ and $\Psi^{AB}_{\phantom{AB}CD}\widetilde{\gamma}^3_{\,A} = \Psi^{AB}_{\phantom{AB}CD}\widetilde{\gamma}^4_{\,A} = 0$ then it follows that the pure spinors $\chi_1$, $\chi_2$, $\widetilde{\gamma}^3$ and $\widetilde{\gamma}^4$ obey to the integrability condition of equations (\ref{Integrab. Pure spinor}) and (\ref{Integrab. Pure spinor NegChiral.})\footnote{Remember that the fact that these four pure spinors obey to the integrability condition is a general property of the type $N$ space-times, see the paragraph after equation (\ref{Integrab. Pure spinor NegChiral.}).}. Indeed, these are respectively the pure spinors associated to the four integrable maximally isotropic distributions described right after equation (\ref{nullFppwave}). Now redefining the frame of self-dual 3-vectors of equation (\ref{3-vector basis}) by making the changes $T_2 \leftrightarrow T_{10}$, $T_3 \leftrightarrow T_{5}$ and $T_4 \leftrightarrow T_8$ we find the following matrix representation for the operator $C^+$:
$$ [C^+]\;=\;  \textrm{Blockdiag} (\left[
                                                    \begin{array}{cc}
                                                      0 & \Theta \\
                                                      0 & 0 \\
                                                    \end{array}
                                                  \right]
 ,\left[
    \begin{array}{cc}
      0 & \widetilde{\Theta} \\
      0 & 0 \\
    \end{array}
  \right]
  ,0,0,0,0,0,0) $$
 From this matrix representation it is seen that all eigenvalues of $C^+$ vanish and that the algebraic type of this operator according to the refined Segre classification is $[|2,2,1,1,1,1,1,1]$.

\section{Conclusions}
The basic tools of spinorial formalism in six dimensions were introduced using index notation and it was shown how to represent low-rank tensors of $SO(6;\mathbb{C})$ in this language. In particular, the spinorial form of bivectors opens the possibility to a new algebraic classification for the bivectors. Also the spinor representation of the Weyl tensor makes clear that this tensor can be interpreted as a map from self-dual 3-vectors to self-dual 3-vectors and this was used to define an algebraic classification for the Weyl tensor. Such map takes particular advantage of the relationship between spinors and isotropic structures. 
It would be interesting, and hopefully valuable, to investigate how the different CMPP types constrain this 3-vector map, just as was done in \cite{Hervik-Discri} with the bivector map.
The approach throughout this article was to work with complex numbers and when necessary choose a real slice according to the space signature. In the spinorial language the imposition of reality conditions was seen to be very simple when the signature is Euclidian, but it will depend on a non-trivial charge conjugation operator in the other signatures. A similar thing happens in four dimensions, in which case the reality conditions are trivial to deal using spinor formalism only when the signature is Lorentzian.

An important result of the present work is the equation (\ref{Integrab. Pure spinor}), expressing the integrability condition for a maximally isotropic distribution in terms of an algebraic condition involving the Weyl ``spinor'' and the pure spinor associated to such distribution. This integrability condition was also proved to be nicely expressed in terms of restrictions on the map of 3-vectors into 3-vectors provided by the Weyl tensor. Particularly, it was proved that in general if a null 3-vector generates an integrable distribution in a Ricci-flat manifold then it is an eigen-3-vector of the Weyl operator. The same result is valid in four dimensions if we put null bivectors instead of null 3-vectors. Finally, the Mariot-Robinson theorem was used to make a link between the existence of a null $3$-form that is harmonic (closed and co-closed) and the existence of a pure spinor obeying the integrability condition.


In four dimensions there is an intimate relationship between the existence of integrable null distributions and the integrability of Einstein's equation, which is somewhat hidden in usual treatments of algebraically special spaces. The 6-dimensional case helps us state that this seems to be the right way of thinking about integrability in higher dimensions. Indeed, throughout this article it was pointed out various similarities between the 4 and the 6-dimensional cases. Thus the present work should have implications in finding exact solutions for 6-dimensional backgrounds, with relevance for string theory compactifications. In most cases, the Ricci-flat condition can be relaxed to ``conformally Ricci-flat'', which includes Einstein spaces. Also, the relationship between algebraically special manifolds and special holonomy is known and has helped generating new supersymmetric backgrounds \cite{Gauntlett:2004hs}. There is however a whole class of solutions which, despite being non-supersymmetric, do not generate a mass gap, and thus are of interest to string compactifications. These are obtained as the Wick rotation of extremal black holes \cite{Guica:2008mu}. Recently a correspondence has been made between the extremality and the algebraic character of the Killing horizon \cite{BGCC:ARQ},  and this better understanding will surely yield new tools to find interesting solutions for compactification studies. Lastly, there is the prospect that a spinorial approach can be helpful on the calculation of scattering amplitudes, as it was in four dimensions \cite{scattering4d-1,scattering4d-2}, by making full use of 6-dimensional Poincar\'{e} invariance.


More work in the direction of the present paper is in progress. The next natural step is to introduce an extension of the Levi-Civita connection in the spinor bundle. This can help to understand better the subtleties of the generalized Goldberg-Sachs theorem introduced by Taghavi-Chabert \cite{HigherGSisotropic2}, in particular this approach can elucidate when the generality assumption on the Weyl tensor is necessary. Moreover the spinorial techniques can help the obtention of a link between integrability of isotropic structures, not necessarily maximal, and restrictions on the optical matrix (shear, twist and expansion). It is also important to further investigate the classification scheme for the Weyl tensor defined in subsection \ref{SubSec 3-vectors MAP} as well as its relation with the optical matrix.

\section*{Acknowledgments}
Carlos Batista thanks to CNPq (Conselho Nacional de Desenvolvimento Cient\'{\i}fico e Tecnol\'{o}gico) for the financial support.

\appendix
\section{Refined Segre Classification}\label{Appendix- Segre}
The Segre classification is a widely known form to classify square matrices using the Jordan canonical form \cite{Segre Type}. In this appendix such classification will be explained and refined.

By means of a similarity transformation every square matrix over the complex field can be put in the so called Jordan canonical form:
\begin{equation}\label{Jordan Form}
  [M]\rightarrow[M]_J\,= [BMB^{-1}]\,=\, \textrm{Blockdiag}(J_1,J_2,\ldots,J_n) \;\;\;;\;\;J_k = \left[
                                                                           \begin{array}{cccc}
                                                                             \lambda_k & 1 & 0 &\ldots\\
                                                                             0 & \lambda_k & 1&  \\
                                                                             \vdots&  & \ddots &1\\
                                                                             0 & \ldots & 0 &   \lambda_k \\
                                                                           \end{array}
                                                                         \right]  \;\;\;\textrm{or} \;\; J_k = \lambda_k.
\end{equation}
Where $\lambda_k$ is a complex number. The matrices $J_k$ are called the Jordan blocks of $M$. Each block $J_k$ admits just one eigenvector and its eigenvalue is $\lambda_k$. The Jordan canonical form of a matrix is unique up to the ordering of the Jordan blocks.

Now given a square matrix $M$, its Segre type is a sequence of numbers inside a square bracket, each number denoting the dimension of a Jordan block of this matrix. The numbers associated to Jordan blocks with the same eigenvalue are put inside a round bracket. For example, the matrix
$$[M]_J\,=\, \left[
    \begin{array}{ccccc}
      \lambda& 1 & 0 & 0 & 0 \\
      0 & \lambda & 0 & 0 & 0 \\
      0 & 0 & \alpha & 0 & 0 \\
      0 & 0 & 0 & \beta & 1 \\
      0 & 0 & 0 & 0 & \beta \\
    \end{array}
  \right]
$$
can have the following Segre types: (i) If $\lambda\neq\alpha\neq\beta\neq\lambda$ then the type is $[2,2,1]$; (ii) If $\lambda=\alpha\neq\beta$ or $\lambda\neq\alpha=\beta$ the Segre type is $[(2,1),2]$; (iii) If $\lambda=\beta\neq\alpha$ then the type is [(2,2),1]; (iv) If $\lambda=\alpha=\beta$ the type of this matrix is $[(2,2,1)]$.

It is useful to refine this classification by making explicit which Jordan blocks have zero eigenvalue. This can be done by positioning, in the Segre classification, the numbers related to the zero eigenvalues on the right of the other numbers, separating by a vertical bar. As an illustration suppose that in the last example we have $\lambda=0$, then the matrix can take the following types in this refined classification: (i) If $\lambda\neq\alpha\neq\beta\neq\lambda$ then the type is $[2,1|2]$; (ii) If $\lambda=\alpha\neq\beta$ the types is $[2|(2,1)]\equiv[2|2,1]$; (ii') If $\lambda\neq\alpha=\beta$ then the refined Segre type is $[(2,1)|2]$; (iii) If $\lambda=\beta\neq\alpha$ then the type is [1|2,2]; (iv) If $\lambda=\alpha=\beta$, all eigenvalues vanish and the type of the above matrix is $[|(2,2,1)]\equiv[|2,2,1]$.

\section{ A Basis for the Spinor Space}\label{Appendix-Spinor Basis}
The space of (Weyl)spinors of $\mathbb{C}^6$ is the vector space where the representation $\textbf{4}$ of $SL(4;\mathbb{C})$ act, this is a four-dimensional complex space. Let $\{\chi_1^{\,A}, \chi_2^{\,A}, \chi_3^{\,A}, \chi_4^{\,A} \}$ be a basis for this space and let us choose the following normalization condition:
\begin{equation}\label{Normalization condition}
    \varepsilon_{ABCD}\,\chi_1^{\,A}\chi_2^{\,B}\chi_3^{\,C}\chi_4^{\,D} \,=\,1
\end{equation}
The dual basis is then given by:
\begin{equation}\label{Dual spinor basis}
    \widetilde{\gamma}^1_{\,A} = \varepsilon_{ABCD}\chi_2^{\,B}\chi_3^{\,C}\chi_4^{\,D}\;\;;\;\;
    \widetilde{\gamma}^2_{\,A} = -\, \varepsilon_{ABCD}\chi_1^{\,B}\chi_3^{\,C}\chi_4^{\,D}\;\;;\;\;
    \widetilde{\gamma}^3_{\,A} = \varepsilon_{ABCD}\chi_1^{\,B}\chi_2^{\,C}\chi_4^{\,D}\;\;;\;\;
    \widetilde{\gamma}^4_{\,A} = -\, \varepsilon_{ABCD}\chi_1^{\,B}\chi_2^{\,C}\chi_3^{\,D}
\end{equation}
The elements of the dual basis transforms according to the representation $\widetilde{\textbf{4}}$ of $SL(4;\mathbb{C})$ and obey to the important relation $\chi_p^{\,A}\widetilde{\gamma}^q_{\,A} = \delta^q_p$. Now we can express a null frame, $\{e_i,e_{j+3}=\theta^j\}$, in terms of the spinor basis:
\begin{equation}\label{Null basis in spinor basis}
    e_1^{\,AB} = \chi_1^{\,[A}\chi_2^{\,B]}\;;\;\; e_2^{\,AB} = \chi_1^{\,[A}\chi_3^{\,B]}\;;\;\; e_3^{\,AB} = \chi_1^{\,[A}\chi_4^{\,B]}\;;\;\; \theta^{1\,AB} = \chi_3^{\,[A}\chi_4^{\,B]}\;;\;\; \theta^{2\,AB} = \chi_4^{\,[A}\chi_2^{\,B]}\;;\;\; \theta^{3\,AB} = \chi_2^{\,[A}\chi_3^{\,B]}
\end{equation}
Lowering the pair of spinorial indices of this basis by means of the relation $V_{AB}\equiv \frac{1}{2} \varepsilon_{ABCD} V^{CD}$ it is seen that:
\begin{equation}\label{Null basis low indices}
    e_{1\,AB} = \widetilde{\gamma}^3_{\,[A}\widetilde{\gamma}^4_{\,B]}\;;\;\;
    e_{2\,AB} = \widetilde{\gamma}^4_{\,[A}\widetilde{\gamma}^2_{\,B]}\;;\;\;
     e_{3\,AB} = \widetilde{\gamma}^2_{\,[A}\widetilde{\gamma}^3_{\,B]}\;;\;\;
      \theta^1_{\,AB} = \widetilde{\gamma}^1_{\,[A}\widetilde{\gamma}^2_{\,B]}\;;\;\;
      \theta^2_{\,AB} = \widetilde{\gamma}^1_{\,[A}\widetilde{\gamma}^3_{\,B]}\;;\;\;
      \theta^3_{\,AB} = \widetilde{\gamma}^1_{\,[A}\widetilde{\gamma}^4_{\,B]}
\end{equation}
Using equations (\ref{Null basis in spinor basis}) and (\ref{Null basis low indices}) it is immediate to verify that the inner products are in accordance with the ones of a null basis, $ e_i^{\,AB}e_{j\,AB} = 0 = \theta^{i\,AB} \theta^j_{\,AB}$ and $e_i^{\,AB}\theta^j_{\,AB}=\frac{1}{2}\delta^j_i$. Note that the specific representation of equation (\ref{e^AB}) is obtained by taking $\chi_p^{\,A}=\delta_p^A$.

As previously seen, in the spinorial formalism a bivector $B_{\mu\nu}=B_{[\mu\nu]}$ is equivalent to $B^A_{\phantom{A}B}$ with $B^A_{\phantom{A}A}=0$.
Defining $(e_a\wedge e_b) \equiv (e_a\otimes e_b - e_b\otimes e_a )$ and using equation (\ref{B^AB from B^AB CD}) it is possible to find the spinorial representation of a bivector basis, the final result is summarized in table \ref{Table Bivectors}. Now given two bivectors, $B'$ and $B$, it follows, by definition, that $B'_{\mu\nu}B^{\mu\nu}=\mathfrak{B}'_{AB\, CD}\mathfrak{B}^{AB\, CD}$, where $\mathfrak{B}$ was defined on subsection \ref{SubSec Precise Ident.}. By means of equation (\ref{B^AB CD}) it is simple matter then to prove that $B'_{\mu\nu}B^{\mu\nu}=-8B'^A_{\phantom{A}B}B^B_{\phantom{B}A}$. Then using this result and equation (\ref{C(B) by spinors}) we arrive at the following important equation:
\begin{equation}\label{Cabcd = B Psi B}
    C_{abcd}= 2^6\,  (e_a\wedge e_b)^A_{\phantom{A}B}\, \Psi^{BC}_{\phantom{BC}AD} \,(e_c\wedge e_d)^D_{\phantom{D}C}\,.
\end{equation}
Table \ref{Table Bivectors} together with the above equation enables us to express the components of the Weyl tensor, $C_{abcd}$, in terms of the components of $\Psi^{AB}_{\phantom{AB}CD}$. For example, $C_{1234}=2^6 (e_1\wedge e_2)^A_{\phantom{A}B}\, \Psi^{BC}_{\phantom{BC}AD} \,(e_3\wedge \theta^1)^D_{\phantom{D}C} = 4 \chi_1^{\,A}\widetilde{\gamma}^4_{\,B} \Psi^{BC}_{\phantom{BC}AD} \chi_4^{\,D}\widetilde{\gamma}^2_{\,C} = 4 \Psi^{42}_{\phantom{42}14}$. The explicit form of all Weyl tensor's components in a null frame in terms of spinors is given in table \ref{Table-WeylComp}.
\begin{table}[!htbp]
\begin{center}
\begin{tabular}{cccc}
  \hline
  $(e_1\wedge e_2)^A_{\phantom{A}B} = -\frac{1}{4} \chi_1^{\,A}\widetilde{\gamma}^4_{\,B}$\;; & $(e_1\wedge e_3)^A_{\phantom{A}B} = \frac{1}{4} \chi_1^{\,A}\widetilde{\gamma}^3_{\,B}$\;; & $(e_1\wedge\theta^2)^A_{\phantom{A}B} = -\frac{1}{4} \chi_2^{\,A}\widetilde{\gamma}^3_{\,B}$\;; & $(e_1\wedge\theta^3)^A_{\phantom{A}B} = -\frac{1}{4} \chi_2^{\,A}\widetilde{\gamma}^4_{\,B}$\;; \\
  $(e_2\wedge e_3)^A_{\phantom{A}B} = -\frac{1}{4} \chi_1^{\,A}\widetilde{\gamma}^2_{\,B}$, & $(e_2\wedge\theta^1)^A_{\phantom{A}B} = -\frac{1}{4} \chi_3^{\,A}\widetilde{\gamma}^2_{\,B}$\;; & $(e_2\wedge\theta^3)^A_{\phantom{A}B} = -\frac{1}{4} \chi_3^{\,A}\widetilde{\gamma}^4_{\,B}$\;; & $(e_3\wedge\theta^1)^A_{\phantom{A}B} = -\frac{1}{4} \chi_4^{\,A}\widetilde{\gamma}^2_{\,B}$\;; \\
  $(e_3\wedge\theta^2)^A_{\phantom{A}B} = -\frac{1}{4} \chi_4^{\,A}\widetilde{\gamma}^3_{\,B}$\;; & $(\theta^1\wedge\theta^2)^A_{\phantom{A}B} = \frac{1}{4} \chi_4^{\,A}\widetilde{\gamma}^1_{\,B}$\;; & $(\theta^1\wedge\theta^3)^A_{\phantom{A}B} = -\frac{1}{4} \chi_3^{\,A}\widetilde{\gamma}^1_{\,B}$\;;  & $(\theta^2\wedge\theta^3)^A_{\phantom{A}B} = \frac{1}{4} \chi_2^{\,A}\widetilde{\gamma}^1_{\,B}$\;;  \\
  \multicolumn{4}{ c }{$ (e_1\wedge\theta^1)^A_{\phantom{A}B} = \frac{1}{8} [- \chi_1^{\,A}\widetilde{\gamma}^1_{\,B} - \chi_2^{\,A}\widetilde{\gamma}^2_{\,B} + \chi_3^{\,A}\widetilde{\gamma}^3_{\,B} + \chi_4^{\,A}\widetilde{\gamma}^4_{\,B}]$\;;} \\
  \multicolumn{4}{ c }{$ (e_2\wedge\theta^2)^A_{\phantom{A}B} = \frac{1}{8} [- \chi_1^{\,A}\widetilde{\gamma}^1_{\,B} + \chi_2^{\,A}\widetilde{\gamma}^2_{\,B} - \chi_3^{\,A}\widetilde{\gamma}^3_{\,B} + \chi_4^{\,A}\widetilde{\gamma}^4_{\,B}]$\;;} \\
  \multicolumn{4}{ c }{$ (e_3\wedge\theta^3)^A_{\phantom{A}B} = \frac{1}{8} [- \chi_1^{\,A}\widetilde{\gamma}^1_{\,B} + \chi_2^{\,A}\widetilde{\gamma}^2_{\,B} + \chi_3^{\,A}\widetilde{\gamma}^3_{\,B} - \chi_4^{\,A}\widetilde{\gamma}^4_{\,B}]$\;;} \\
\end{tabular}
\newline
\newline
\begin{tabular}{cc}
$\chi_1^{\,A}\widetilde{\gamma}^1_{\,B}  = 2[\frac{1}{8}\delta^A_B - (e_1\wedge\theta^1)^A_{\phantom{A}B} - (e_2\wedge\theta^2)^A_{\phantom{A}B} - (e_3\wedge\theta^3)^A_{\phantom{A}B}]$ \;;& $\chi_2^{\,A}\widetilde{\gamma}^2_{\,B}  = 2[\frac{1}{8}\delta^A_B - (e_1\wedge\theta^1)^A_{\phantom{A}B} + (e_2\wedge\theta^2)^A_{\phantom{A}B} + (e_3\wedge\theta^3)^A_{\phantom{A}B}]$\;; \\
$\chi_3^{\,A}\widetilde{\gamma}^3_{\,B}  = 2[\frac{1}{8}\delta^A_B + (e_1\wedge\theta^1)^A_{\phantom{A}B} - (e_2\wedge\theta^2)^A_{\phantom{A}B} + (e_3\wedge\theta^3)^A_{\phantom{A}B}]$\;; &
$\chi_4^{\,A}\widetilde{\gamma}^4_{\,B}  = 2[\frac{1}{8}\delta^A_B + (e_1\wedge\theta^1)^A_{\phantom{A}B} + (e_2\wedge\theta^2)^A_{\phantom{A}B} - (e_3\wedge\theta^3)^A_{\phantom{A}B}]$\;. \\ \hline
\end{tabular}
\caption{The spinorial representation of a bivector basis. The last two lines display the inverse relation of the 3 short lines at the center.}\label{Table Bivectors}
\end{center}
\end{table}

\begin{table}[!htbp]
\begin{center}
\begin{tabular}{ccccc}
  \hline
  $C_{1212}=4\Psi^{44}_{\phantom{44}11}$ \; ; \;$C_{1213}=-4\Psi^{34}_{\phantom{34}11}$ \; ; \; $C_{1215}=4\Psi^{34}_{\phantom{34}12}$ \; ; \; $C_{1216}=4\Psi^{44}_{\phantom{44}12}$ \; ; \; $C_{1313}=4\Psi^{33}_{\phantom{33}11}$\\
    $C_{1315}=-4\Psi^{33}_{\phantom{33}12}$ \; ; \; $C_{1316}=-4\Psi^{34}_{\phantom{34}12}$ \; ; \;$C_{1515}=4\Psi^{33}_{\phantom{33}22}$ \; ; \; $C_{1516}=4\Psi^{34}_{\phantom{34}32}$\; ; \; $C_{1616}=4\Psi^{44}_{\phantom{44}22}$\\
 \end{tabular}
  \begin{tabular}{ccc}
  $C_{1223}=4\Psi^{24}_{\phantom{24}11}$ \; ; \;$C_{1225}=4(\Psi^{14}_{\phantom{14}11}+\Psi^{34}_{\phantom{34}31})=-4(\Psi^{44}_{\phantom{44}41}+\Psi^{24}_{\phantom{24}21})$ \; ; \;$C_{1226}=4\Psi^{44}_{\phantom{44}13}$ \\
  $C_{1235}=4\Psi^{34}_{\phantom{34}14}$ \; ; \;$C_{1236}=4(\Psi^{14}_{\phantom{14}11}+\Psi^{44}_{\phantom{44}41})=-4(\Psi^{24}_{\phantom{24}21}+\Psi^{34}_{\phantom{34}31})$ \; ; \;$C_{1256}=-4\Psi^{14}_{\phantom{14}12}$ \\
  $C_{1323}=-4\Psi^{23}_{\phantom{23}11}$ \; ; \;$C_{1325}=4(\Psi^{23}_{\phantom{23}21}+\Psi^{43}_{\phantom{43}41})=-4(\Psi^{13}_{\phantom{13}11}+\Psi^{33}_{\phantom{33}31})$ \; ; \;$C_{1326}=-4\Psi^{43}_{\phantom{43}13}$ \\
  $C_{1335}=-4\Psi^{33}_{\phantom{33}14}$ \; ; \;$C_{1336}=4(\Psi^{23}_{\phantom{23}21}+\Psi^{33}_{\phantom{33}31})=-4(\Psi^{13}_{\phantom{13}11}+\Psi^{43}_{\phantom{43}41})$ \; ; \;$C_{1356}=4\Psi^{13}_{\phantom{13}12}$ \\
  $C_{1523}=4\Psi^{32}_{\phantom{32}12}$ \; ; \;$C_{1525}=4(\Psi^{13}_{\phantom{13}12}+\Psi^{33}_{\phantom{33}32})=-4(\Psi^{23}_{\phantom{23}22}+\Psi^{43}_{\phantom{43}42})$ \; ; \;$C_{1526}=4\Psi^{34}_{\phantom{34}23}$ \\
  $C_{1535}=4\Psi^{33}_{\phantom{33}24}$ \; ; \;$C_{1536}=4(\Psi^{13}_{\phantom{13}12}+\Psi^{43}_{\phantom{43}42})=-4(\Psi^{23}_{\phantom{23}22}+\Psi^{33}_{\phantom{33}32})$ \; ; \;$C_{1556}=-4\Psi^{13}_{\phantom{13}22}$ \\
  $C_{1623}=4\Psi^{24}_{\phantom{24}21}$ \; ; \;$C_{1625}=4(\Psi^{14}_{\phantom{14}12}+\Psi^{34}_{\phantom{34}32})=-4(\Psi^{24}_{\phantom{24}22}+\Psi^{44}_{\phantom{44}42})$ \; ; \;$C_{1626}=4\Psi^{44}_{\phantom{44}23}$ \\
  $C_{1635}=4\Psi^{34}_{\phantom{34}24}$ \; ; \;$C_{1636}=4(\Psi^{14}_{\phantom{14}12}+\Psi^{44}_{\phantom{44}42})=-4(\Psi^{24}_{\phantom{24}22}+\Psi^{34}_{\phantom{34}32})$ \; ; \;$C_{1656}=-4\Psi^{14}_{\phantom{14}22}$ \\
  \end{tabular}
  \begin{tabular}{cc}
  $C_{1412}=4(\Psi^{14}_{\phantom{14}11}+\Psi^{24}_{\phantom{24}21})=-4(\Psi^{34}_{\phantom{34}31}+\Psi^{44}_{\phantom{44}41})$ \; ; \; $C_{1413}=4(\Psi^{33}_{\phantom{33}31}+\Psi^{43}_{\phantom{43}41})=-4(\Psi^{13}_{\phantom{13}11}+\Psi^{23}_{\phantom{23}21})$ \\
  $C_{1415}=4(\Psi^{13}_{\phantom{13}12}+\Psi^{23}_{\phantom{23}22})=-4(\Psi^{33}_{\phantom{33}32}+\Psi^{43}_{\phantom{43}42})$ \; ; \; $C_{1416}=4(\Psi^{14}_{\phantom{14}12}+\Psi^{24}_{\phantom{24}22})=-4(\Psi^{34}_{\phantom{34}32}+\Psi^{44}_{\phantom{44}42})$\\
  \end{tabular}
  \begin{tabular}{cc}
  $C_{1414}=4(\Psi^{11}_{\phantom{11}11}+\Psi^{22}_{\phantom{22}22}+2\Psi^{12}_{\phantom{12}12})=
  4(\Psi^{33}_{\phantom{33}33}+\Psi^{44}_{\phantom{44}44}+2\Psi^{34}_{\phantom{34}34})$\; ; \; $C_{1425}=4(\Psi^{23}_{\phantom{23}23}-\Psi^{14}_{\phantom{14}14})$\\
  $C_{2525}=4(\Psi^{11}_{\phantom{11}11}+\Psi^{33}_{\phantom{33}33}+2\Psi^{13}_{\phantom{13}13})=
  4(\Psi^{22}_{\phantom{22}22}+\Psi^{44}_{\phantom{44}44}+2\Psi^{24}_{\phantom{24}24})$\; ; \; $C_{1436}=4(\Psi^{24}_{\phantom{24}24}-\Psi^{13}_{\phantom{13}13})$\\
  $C_{3636}=4(\Psi^{11}_{\phantom{11}11}+\Psi^{44}_{\phantom{44}44}+2\Psi^{14}_{\phantom{14}14})=
  4(\Psi^{22}_{\phantom{22}22}+\Psi^{33}_{\phantom{33}33}+2\Psi^{23}_{\phantom{23}23})$\; ; \; $C_{2536}=4(\Psi^{34}_{\phantom{34}34}-\Psi^{12}_{\phantom{12}12})$\\
  \end{tabular}
  \begin{tabular}{cc}
  $C_{1423}=4(\Psi^{12}_{\phantom{12}11}+\Psi^{22}_{\phantom{22}21})=-4(\Psi^{32}_{\phantom{32}31}+\Psi^{42}_{\phantom{42}41})$ \; ; \; $C_{1426}=4(\Psi^{14}_{\phantom{14}13}+\Psi^{24}_{\phantom{24}23})=-4(\Psi^{34}_{\phantom{34}33}+\Psi^{44}_{\phantom{44}43})$ \\
  $C_{1435}=4(\Psi^{13}_{\phantom{13}14}+\Psi^{23}_{\phantom{23}24})=-4(\Psi^{33}_{\phantom{33}34}+\Psi^{43}_{\phantom{43}44})$ \; ; \; $C_{1456}=4(\Psi^{31}_{\phantom{31}32}+\Psi^{41}_{\phantom{41}42})=-4(\Psi^{11}_{\phantom{11}12}+\Psi^{21}_{\phantom{21}22})$ \\
  $C_{2523}=4(\Psi^{12}_{\phantom{12}11}+\Psi^{23}_{\phantom{23}13})=-4(\Psi^{22}_{\phantom{22}12}+\Psi^{24}_{\phantom{24}14})$ \; ; \; $C_{3623}=4(\Psi^{12}_{\phantom{12}11}+\Psi^{24}_{\phantom{24}14})=-4(\Psi^{22}_{\phantom{22}12}+\Psi^{23}_{\phantom{23}13})$ \\
  $C_{2526}=4(\Psi^{14}_{\phantom{14}13}+\Psi^{34}_{\phantom{34}33})=-4(\Psi^{24}_{\phantom{24}23}+\Psi^{44}_{\phantom{44}34})$ \; ; \; $C_{3626}=4(\Psi^{14}_{\phantom{14}13}+\Psi^{44}_{\phantom{44}34})=-4(\Psi^{24}_{\phantom{24}23}+\Psi^{34}_{\phantom{34}33})$ \\
  $C_{2535}=4(\Psi^{13}_{\phantom{13}14}+\Psi^{33}_{\phantom{33}34})=-4(\Psi^{23}_{\phantom{23}24}+\Psi^{34}_{\phantom{34}44})$ \; ; \; $C_{3635}=4(\Psi^{13}_{\phantom{13}14}+\Psi^{34}_{\phantom{34}44})=-4(\Psi^{23}_{\phantom{23}24}+\Psi^{33}_{\phantom{33}34})$ \\
  $C_{2556}=4(\Psi^{12}_{\phantom{12}22}+\Psi^{14}_{\phantom{14}24})=-4(\Psi^{11}_{\phantom{11}12}+\Psi^{13}_{\phantom{13}23})$ \; ; \; $C_{3656}=4(\Psi^{12}_{\phantom{12}22}+\Psi^{13}_{\phantom{13}23})=-4(\Psi^{11}_{\phantom{11}12}+\Psi^{14}_{\phantom{14}24})$ \\
  \end{tabular}
  \begin{tabular}{ccccc}
  $C_{2323}=4\Psi^{22}_{\phantom{22}11}$ \; ; \;$C_{2326}=4\Psi^{24}_{\phantom{24}13}$ \; ; \; $C_{2335}=4\Psi^{23}_{\phantom{23}14}$ \; ; \; $C_{2356}=-4\Psi^{12}_{\phantom{12}12}$ \; ; \; $C_{5656}=4\Psi^{11}_{\phantom{11}22}$\\
  $C_{2626}=4\Psi^{44}_{\phantom{44}33}$ \; ; \;$C_{2635}=4\Psi^{34}_{\phantom{34}34}$ \; ; \; $C_{2656}=-4\Psi^{14}_{\phantom{14}23}$ \; ; \; $C_{3535}=4\Psi^{33}_{\phantom{33}44}$ \; ; \; $C_{3556}=-4\Psi^{13}_{\phantom{13}24}$ \\

  $C_{1242}=-4\Psi^{24}_{\phantom{24}13}$ \; ; \;$C_{1243}=-4\Psi^{24}_{\phantom{24}14}$ \; ; \; $C_{1245}=-4\Psi^{14}_{\phantom{14}14}$ \; ; \; $C_{1246}=4\Psi^{14}_{\phantom{14}13}$ \; ; \; $C_{1342}=4\Psi^{23}_{\phantom{23}13}$\\

  $C_{1343}=4\Psi^{23}_{\phantom{23}14}$ \; ; \;$C_{1345}=4\Psi^{13}_{\phantom{13}14}$ \; ; \; $C_{1346}=-4\Psi^{13}_{\phantom{13}13}$ \; ; \; $C_{1542}=-4\Psi^{23}_{\phantom{23}23}$ \; ; \; $C_{1543}=-4\Psi^{23}_{\phantom{23}24}$\\
  \end{tabular}
  \begin{tabular}{cccccc}
  $C_{1545}=-4\Psi^{13}_{\phantom{13}24}$ \; ; \;$C_{1546}=4\Psi^{13}_{\phantom{13}23}$ \; ; \; $C_{1642}=-4\Psi^{24}_{\phantom{24}23}$ \; ; \; $C_{1643}=-4\Psi^{24}_{\phantom{24}24}$ \; ; \; $C_{1645}=-4\Psi^{14}_{\phantom{14}24}$ \; ; \; $C_{1646}=4\Psi^{14}_{\phantom{14}23}$\\
    \hline
  \end{tabular}
\caption{This table displays the relation between Weyl tensor's components in a null frame and its spinorial equivalent. The missing components of the Weyl tensor can be obtained by making the changes $1\leftrightarrow4$, $2\leftrightarrow5$ and $3\leftrightarrow6$ on the vectorial indices while swapping the upper and the lower indices of $\Psi$. The first two rows of the above table contain the components of the Weyl tensor with boost weight $b=2$, the next ten rows present the components with $b=1$, the other rows have the components with zero boost weight.}\label{Table-WeylComp}
\end{center}
\end{table}

 With the tolls introduced in this appendix  it is simple matter to show that $\chi_1^{\,D}\varepsilon_{DABC}=6\widetilde{\gamma}^2_{\,[A}\widetilde{\gamma}^3_{\,B}\widetilde{\gamma}^4_{\,C]}$. Using this relation it is then possible to prove, after some algebra, the following equality:
\begin{multline}\label{EE Psi XXX}
  \frac{1}{4} (\,\varepsilon_{AEFG}\,\varepsilon_{BHIJ}\,\Psi^{GJ}_{\phantom{GJ}CD}\,)\,\chi_1^{\,A}\, \chi_1^{\,B} \, \chi_1^{\,C} \,=\,  \Psi^{44}_{\phantom{44}1D} \widetilde{\gamma}^2_{\,[E}\widetilde{\gamma}^3_{\,F]}\widetilde{\gamma}^2_{\,[H}\widetilde{\gamma}^3_{\,I]} \,+\,  \Psi^{43}_{\phantom{43}1D} (\widetilde{\gamma}^2_{\,[E}\widetilde{\gamma}^3_{\,F]}\widetilde{\gamma}^4_{\,[H}\widetilde{\gamma}^2_{\,I]} + \widetilde{\gamma}^4_{\,[E}\widetilde{\gamma}^2_{\,F]}\widetilde{\gamma}^2_{\,[H}\widetilde{\gamma}^3_{\,I]}) \,+\,  \\   +\,
  \Psi^{42}_{\phantom{42}1D} (\widetilde{\gamma}^2_{\,[E}\widetilde{\gamma}^3_{\,F]}\widetilde{\gamma}^3_{\,[H}\widetilde{\gamma}^4_{\,I]} + \widetilde{\gamma}^3_{\,[E}\widetilde{\gamma}^4_{\,F]}\widetilde{\gamma}^2_{\,[H}\widetilde{\gamma}^3_{\,I]})  \,+\, \Psi^{33}_{\phantom{33}1D} \widetilde{\gamma}^2_{\,[E}\widetilde{\gamma}^4_{\,F]}\widetilde{\gamma}^2_{\,[H}\widetilde{\gamma}^4_{\,I]}\,+\, \\  +\, \Psi^{32}_{\phantom{42}1D} (\widetilde{\gamma}^2_{\,[E}\widetilde{\gamma}^4_{\,F]}\widetilde{\gamma}^4_{\,[H}\widetilde{\gamma}^3_{\,I]} + \widetilde{\gamma}^4_{\,[E}\widetilde{\gamma}^3_{\,F]}\widetilde{\gamma}^2_{\,[H}\widetilde{\gamma}^4_{\,I]}) \,+\, \Psi^{22}_{\phantom{22}1D} \widetilde{\gamma}^3_{\,[E}\widetilde{\gamma}^4_{\,F]}\widetilde{\gamma}^3_{\,[H}\widetilde{\gamma}^4_{\,I]}.
 \end{multline}

Now let us define a basis for the space of 3-vectors:
\begin{multline}\label{3-vector basis}
   T_1^{\;AB}= \chi_1^{\,A}\chi_1^{\,B} \; ; \;  T_2^{\;AB}= \sqrt{2}\,\chi_1^{\,(A}\chi_2^{\,B)} \; ; \; T_3^{\;AB}= \sqrt{2}\,\chi_1^{\,(A}\chi_3^{\,B)} \; ; \; T_4^{\;AB}= \sqrt{2}\,\chi_1^{\,(A}\chi_4^{\,B)} \; ; \; T_5^{\;AB}=  \chi_2^{\,A}\chi_2^{\,B} \; ; \; T_6^{\;AB}= \sqrt{2}\,\chi_2^{\,(A}\chi_3^{\,B)}\\
   T_7^{\;AB}= \sqrt{2}\,\chi_2^{\,(A}\chi_4^{\,B)} \; ; \;T_8^{\;AB}= \chi_3^{\,A}\chi_3^{\,B}   \; ; \;T_9^{\;AB}=  \sqrt{2}\,\chi_3^{\,(A}\chi_4^{\,B)} \; ; \;T_{10}^{\;AB}= \chi_4^{\,A}\chi_4^{\,B}
   \quad \quad \quad \quad \\
   \widetilde{T}^1_{\;AB}= \widetilde{\gamma}^1_{\,A}\widetilde{\gamma}^1_{\,B} \; ; \; \widetilde{T}^2_{\;AB}= \sqrt{2} \widetilde{\gamma}^1_{\,(A}\widetilde{\gamma}^2_{\,B)}\; ; \; \widetilde{T}^3_{\;AB}= \sqrt{2} \widetilde{\gamma}^1_{\,(A}\widetilde{\gamma}^3_{\,B)}\; ; \; \widetilde{T}^4_{\;AB}= \sqrt{2} \widetilde{\gamma}^1_{\,(A}\widetilde{\gamma}^4_{\,B)}  \; ; \; \widetilde{T}^5_{\;AB}= \widetilde{\gamma}^2_{\,A}\widetilde{\gamma}^2_{\,B} \;;\;
   \widetilde{T}^6_{\;AB}= \sqrt{2} \widetilde{\gamma}^2_{\,(A}\widetilde{\gamma}^3_{\,B)} \;\\
   \widetilde{T}^7_{\;AB}= \sqrt{2} \widetilde{\gamma}^2_{\,(A}\widetilde{\gamma}^4_{\,B)}  \; ; \;
   \widetilde{T}^8_{\;AB}=  \widetilde{\gamma}^3_{\,A}\widetilde{\gamma}^3_{\,B}  \; ; \;
   \widetilde{T}^9_{\;AB}=\sqrt{2} \widetilde{\gamma}^3_{\,(A}\widetilde{\gamma}^4_{\,B)} \; ; \;
   \widetilde{T}^{10}_{\;AB}= \widetilde{\gamma}^4_{\,A}\widetilde{\gamma}^4_{\,B} \quad \quad \quad \quad \quad \quad \end{multline}
   Where by $T_{r}^{\;AB}$ it is meant the 3-vector $(T_{r}^{\;AB},0)$, while $\widetilde{T}^r_{\;AB}$ means $(0,\widetilde{T}^r_{\;AB})$. The ten 3-vectors $\{T_r\}$ form a basis for the space of self-dual 3-vectors, while $\{\widetilde{T}^r\}$ is a basis for the space of anti-self-dual 3-vectors. It is easily seen that $T_{r}^{\;AB}\widetilde{T}^s_{\;AB} = \delta^s_r$.

\section{Clifford Algebra}\label{Appendix - Clifford Alg.}
The space of the Dirac spinors, $S$, is the 8-dimensional space spanned by the spinors $\{\chi_p^{\,A},\widetilde{\gamma}^q_{\,B}\}$. The 4-dimensional subspace generated by $\{\chi_p^{\,A}\}$ is the space of Weyl spinors of positive chirality, $S^+$, while $\{\widetilde{\gamma}^q_{\,B}\}$ spans the subspace of negative chirality Weyl spinors, $S^-$. So a Dirac spinor can be written as $\widehat{\psi}=\psi^A + \widetilde{\psi}_A$, where $\psi^A$ pertains to $S^+$ and $\widetilde{\psi}_A$ belongs to $S^-$. The inner product of two Dirac spinors is defined by:
\begin{equation}\label{Inner prod spinors}
    (\widehat{\psi}_1,\widehat{\psi}_2) = \psi_1^{\,A} \,\widetilde{\psi}_{2\,A} - \psi_2^{\,A} \, \widetilde{\psi}_{1\,A}.
\end{equation}
Note that this inner product is skew-symmetric.

In the Clifford algebra formalism the vectors are seen as linear operators that act on the space $S$, $\textbf{e}_{a}:S \rightarrow S$. These maps must obey to the following relation:
\begin{equation}\label{Clifford product}
    \textbf{e}_{a}\textbf{e}_{b} + \textbf{e}_{b}\textbf{e}_{a} = 2g(e_a,e_b)\,\textbf{1} = 2g_{ab}\,\textbf{1}.
\end{equation}
Where $\textbf{1}$ is the identity operator on $S$. An explicit expression for these operators in terms of (\ref{Null basis in spinor basis}) and (\ref{Null basis low indices}) is given by:
\begin{equation}\label{Clifford algebra e_a}
    \textbf{e}_a = 2(e_a^{\,AB} - e_{a\,AB})\,:\;\;\;\;\; \textbf{e}_a(\widehat{\psi}) = 2 e_a^{\,AB}\widetilde{\psi}_B - 2 e_{a\,AB} \psi^B.
\end{equation}



\begin{thebibliography}{58}
\bibitem{WardWells} R. S. Ward and R. O. Wells Jr., \textit{Twistor Geometry and Field Theory}, Cambridge University Press (1990).

\bibitem{scattering4d-1} R. Britto, F. Cachazo, B. Feng and E. Witten, \textit{Direct proof of tree-level recursion relation in Yang-Mills theory},  Physical  Review Letters {\bf 94} (2005), 181602. Available at arXiv:hep-th/0501052

\bibitem{HuggettTod} S. A. Huggett and K. P. Tod, \textit{An Introduction to Twistor Theory}, Cambridge University Press (1994).



\bibitem{PenroseSpinor} R. Penrose, \textit{A spinor approach to General Relativity}, Annals of Physics \textbf{10} (1960), 171.\\
R. Penrose, \textit{Zero rest-mass fields including gravitation: asymptotic behaviour}, Proceedings of the Royal Society A \textbf{284} (1965), 159.
\bibitem{Penrose} R. Penrose and W. Rindler, \textit{Spinors and space-time} vol.\textbf{1} and\textbf{ 2}, Cambridge
    University Press (1984 and 1986).
\bibitem{AdvancedGR} J. Stewart, \textit{Advanced General Relativity}, Cambridge University Press (1991).
\bibitem{Spinors in 6D} C. S\"{a}mann and M. Wolf, \textit{On twistors and conformal field theories from six dimensions}, (2011). Journal of Mathematical Physics \textbf{54} (2013), 013507. Available at arXiv:1111.2539 \\
    S. Weinberg, \textit{Six-dimensional methods for four-dimensional conformal field theories}, (2010). Physical Review D \textbf{82} (2010), 045031. Available at arXiv:1006.3480

\bibitem{Kerr-Robinson} L. P. Hughston and L. J. Mason, \textit{A generalised Kerr-Robinson theorem}, Classical and Quantum Gravity \textbf{5} (1988), 275.


\bibitem{Kerr} R. P. Kerr, \textit{Gravitational field of a spinning mass as an example of algebraically special metrics}, Physical Review Letters \textbf{11} (1963), 237.
\bibitem{typeD - Kinnersley} W. Kinnersley, \textit{Type D Vacuum Metrics}, Journal of Mathematical Physics \textbf{10} (1969), 1195.

\bibitem{Goldberg-Sachs} J. Goldberg and R. Sachs, \textit{A theorem on Petrov types}, General Relativity and Gravitation \textbf{41} (2009), 433.  This is a republication of the original 1962 paper.
\bibitem{Plebanski2} J. F. Pleba\'{n}ski and S. Hacyan, \textit{Null geodesic surfaces and Goldberg-Sachs theorem in
    complex Riemannian spaces}, Journal of Mathematical Physics \textbf{16} (1975), 2403.
\bibitem{art2} C. Batista, \textit{A Generalization of the Goldberg-Sachs Theorem and its Consequences}, accepted for publication in General Relativity and Gravitation (2013), DOI:10.1007/s10714-013-1539-4. Available at arXiv:1205.4666.


\bibitem{5D classification} P. De Smet, \textit{Black holes on cylinders are not algebraically special}, Classical and Quantum Gravity \textbf{19} (2002), 4877. Available at arXiv:hep-th/0206106 \\
    M. Godazgar, \textit{Spinor classification of the Weyl tensor in five dimensions}, Classical and Quantum Gravity \textbf{27} (2010), 245013. Available at arXiv:1008.2955

\bibitem{CMPP} A. Coley, R. Milson, V. Pravda and A. Pravdov\'{a}, \textit{Classification of the Weyl Tensor in Higher Dimensions}, Classical and Quantum Gravity \textbf{21} (2004), L-35. Available at arXiv:gr-qc/0401008
\bibitem{Coley_Review} A. Coley, \textit{Classification of the Weyl tensor in higher dimensions and applications}, Classical and Quantum Gravity \textbf{25} (2008), 033001. Available at arXiv:0710.1598





\bibitem{Durkee Reall}  M. Durkee and H. S. Reall, \textit{A higher-dimensional generalization of the geodesic
part of the Goldberg-Sachs theorem}, Classical and Quantum Gravity \textbf{26} (2009), 245005. Available at arXiv:0908.2771

\bibitem{M. Orataggio- GSII} M. Ortaggio, V. Pravda and A. Pravdov\'{a}, \textit{On the Goldberg-Sachs theorem in higher dimensions in the non-twisting case}, Classical and Quantum Gravity \textbf{30} (2013), 075016.  Available at arXiv:1211.2660

\bibitem{M. Ortaggio-GS theorem} M. Ortaggio \textit{et. al.}, \textit{On a five dimensional version of the Goldberg-Sachs theorem}, Classical and Quantum Gravity \textbf{29} (2012), 205002. Available at arXiv:1205.1119

\bibitem{M. Ortaggio-Robinson-Trautman} M. Ortaggio, \textit{Higher dimensional spacetimes with a geodesic shear-free, twistfree and expanding null congruence}, Proceedings of the XVII SIGRAV Conference (2006). Available at arXiv:gr-qc/0701036
\bibitem{HigherGSisotropic1} A. Taghavi-Chabert, \textit{Optical structures, algebraically special spacetimes and the
    Goldberg-Sachs theorem in five dimensions}, Classical and Quantum Gravity \textbf{28} (2011), 145010. Available at arXiv:1011.6168
\bibitem{HigherGSisotropic2} A. Taghavi-Chabert, \textit{The complex Goldberg-Sachs theorem in higher dimensions},
    Journal of Geometry and Physics \textbf{62} (2012), 981. Available at arXiv:1107.2283



\bibitem{Lounesto} P. Lounesto,\textit{ Clifford algebras and spinors},  Cambridge University Press (2001).
\bibitem{Slansky} R. Slansky, \textit{Group theory for unified model building}, Physics Reports \textbf{79} (1981), 1.
\bibitem{Trautman} W. Kopczynski and A. Trautman, \textit{Simple spinors and real structures}, Journal of Mathematical Physics \textbf{33} (1992), 550.
\bibitem{art1} C. Batista, \textit{Weyl tensor classifcation in four-dimensional manifolds of all signatures}, General Relativity and Gravitation \textbf{45} (2013), 785. Available at arXiv:1204.5133.


\bibitem{Robinson Manifolds} P. Nurowski and A. Trautman, \textit{Robinson manifolds as the Lorentzian analogs of
    Hermite Manifolds}, Differential Geometry and its Applications \textbf{17}(2002), 175. Available at arXiv:math/0201266

\bibitem{Pravda type D} V. Pravda \textit{et. al.}, \textit{Type D Einstein spacetimes in higher dimensions}, Classical and Quantum Gravity \textbf{24} (2007), 4407.  Available at arXiv:0704.0435

\bibitem{Reall - Review} H. S. Reall, \textit{Algebraically special solutions in higher dimensions}, Black holes in higher dimensions, Cambridge University Press (2012).  Available at arXiv:1105.4057

\bibitem{Plebanski 1} J. Pleba\'{n}ski, \textit{Some solutions of complex Einstein equations}, Journal of Mathematical Physics \textbf{16} (1975), 2395.

\bibitem{Petrov} A. Z. Petrov, \textit{The classification of spaces definig gravitational fields}, General Relativity
    and Gravitation \textbf{32} (2000), 1665. This is a translated republication of the original 1954 paper.

\bibitem{Plebanski3} J. Pleba\'{n}ski and I. Robinson, \textit{Left-degenerate vacuum metrics}, Physical Review Letters \textbf{37} (1976), 493.
\bibitem{Bivector_Coley} A. Coley and S. Hervik, \textit{Higher dimensional bivectors and classification of the Weyl operator}, Classical and Quantum Gravity \textbf{27} (2010), 015002. Availabe at arXiv:0909.1160







\bibitem{CMPP-Bel} R. Milson \textit{et. al.}, \textit{Alignment and algebraically special tensors in Lorentzian geometry}, International Journal of Geometric Methods in Modern Physics \textbf{2} (2005), 41. Available at arXiv:gr-qc/0401010\\
    M. Ortaggio, \textit{Bel-Debever criteria for the classification of the Weyl tensors in higher dimensions}, Classical and Quantum Gravity \textbf{26} (2009), 195015. Available at arXiv:gr-qc/0906.3818

\bibitem{Reall - continuous WAND} M. Godazgar and H.S. Reall, \textit{Algebraically special axisymmetric solutions of the higher-dimensional vacuum Einstein equation}, Classical and Quantum Gravity \textbf{26} (2009), 165009. Available at arXiv:0904.4368
\bibitem{ColeyPSEUD} S. Hervik and A. Coley, \textit{On the algebraic classification of pseudo-Riemannian spaces}, International Journal of Geometric Methods in Modern Physics \textbf{8} (2011), 1679. Available at arXiv:1008.3021
\bibitem{Hervik-VSI} S. Hervik and A. Coley, \textit{Pseudo-Riemannian VSI spaces}, Classical and Quantum Gravity \textbf{28} (2011), 015008. Available at arXiv:1008.2838






\bibitem{Mason-Chabert-KillingYano} L. Mason and A. Taghavi-Chabert, \textit{Killing-Yano tensors and multi-Hermitean structures}, Journal of Geometry and Physics \textbf{60} (2010), 907. Available at arXiv:0805.3756
\bibitem{FrolovMyers5D} V. Frolov and D. Stojkovi\'{c}, \textit{Particle and light motion in a space-time of a five-dimensional black hole},
Physical Review D \textbf{68} (2003), 064011. Available at arXiv:gr-qc/0301016

\bibitem{Robinson}  I. Robinson, \textit{Null Electromagnetic Fields}, Journal of Mathematical Physics \textbf{2} (1961), 290.
\bibitem{McIntoshII} G. S. Hall, M. Hickman and C. McIntosh,  \textit{Complex relativity and real solutions II: Classification of complex bivectors and metric classes}, General Relativity and Gravitation \textbf{17} (1985), 475.

\bibitem{art4} C. Batista, \textit{On the Weyl tensor classification in all dimensions and its relation with integrability properties}, to appear soon.

\bibitem{Kerr-Schild} M. Ortaggio, V. Pravda and A. Pravdov\'{a}, \textit{Higher dimensional Kerr-Schild spacetimes}, Classical and Quantum Gravity \textbf{26} (2009), 025008. Available at arXiv:0808.2165
\bibitem{Hervik-Discri} A. Coley and S. Hervik, \textit{Discriminating the Weyl type in higher dimensions using scalar curvature invariants}, General Relativity
    and Gravitation \textbf{43} (2011), 2199.\\
    A. Coley and S. Hervik, \textit{Algebraic classification of spacetimes using discriminating scalar curvature invariants}, (2010). Available at arXiv:1011.2175







\bibitem{Gauntlett:2004hs}
  J.~P.~Gauntlett, D.~Martelli, J.~Sparks and D.~Waldram,
  {\it Supersymmetric AdS backgrounds in string and M-theory} (2004),
  Available at arXiv:hep-th/0411194.

\bibitem{Guica:2008mu}
  M.~Guica, T.~Hartman, W.~Song and A.~Strominger,
  {\it The Kerr/CFT correspondence,}
  Physical  Review D {\bf 80} (2009), 124008.
  Available at arXiv:0809.4266

\bibitem{BGCC:ARQ}
  B. Carneiro da Cunha and A. R. de Queiroz, to appear.


\bibitem{scattering4d-2} C. F. Berger \textit{et.al}, \textit{One-Loop calculations with BlackHat}, Nuclear Physics Proceedings Supplements {\bf 183} (2008), 313. Available at arXiv:0807.3705

\bibitem{Segre Type} H. Stephani \textit{et. al.}, \textit{Exact solutions of Einstein's field equations}, Cambridge University Press (2009).\\
  J. Santos \textit{et. al.}, \textit{Classification of second order symmetric tensors in 5-dimensional Kaluza-Klein-type theories}, Journal of Mathematical Physics \textbf{36} (1995), 3074. Available at arXiv:gr-qc/9506031



\end{thebibliography}
\end{document}